\documentclass[preprint,12pt,3p]{elsarticle}

\usepackage{amssymb}
\usepackage{lineno,hyperref}
\usepackage{amsmath}
\usepackage{bm}
\usepackage{subcaption}
\usepackage{graphicx}
\usepackage{svg}
\usepackage{rotating}
\usepackage{lscape}
\usepackage{array,multirow}
\usepackage[title]{appendix}
\usepackage{float}
\usepackage{stmaryrd}
\usepackage{textcomp}
\usepackage{algorithm}
\usepackage{algpseudocode}
\usepackage{amsthm}
\usepackage{listings}
\usepackage{xcolor}

\DeclareMathOperator{\tr}{tr}

\theoremstyle{definition}

\newtheorem{remark}{Remark}

\usepackage{siunitx}
\usepackage{symbols}
\usepackage{pifont}

\usepackage{lineno}

\makeatletter
\def\BState{\State\hskip-\ALG@thistlm}
\makeatother

\journal{Elsevier}

\begin{document} 

\begin{frontmatter}

\title{A Locally Conservative Mixed Finite Element Framework for Coupled Hydro-Mechanical-Chemical Processes in Heterogeneous Porous Media}

\author[label1,label2]{T. Kadeethum\corref{cor1}}
\address[label1]{Technical University of Denmark, Denmark}
\address[label2]{Cornell University, New York, USA}

\cortext[cor1]{corresponding author}

\ead{teekad@dtu.dk}

\author[label3]{S. Lee}
\address[label3]{Florida State University, Florida, USA}
\ead{lee@math.fsu.edu}

\author[label4]{F. Ballarin}
\address[label4]{mathLab, Mathematics Area, SISSA, Italy}
\ead{francesco.ballarin@sissa.it}

\author[label5]{J. Choo}
\address[label5]{The University of Hong Kong, Hong Kong}
\ead{jchoo@hku.hk}

\author[label1]{H.M. Nick}
\ead{hamid@dtu.dk}

\fntext[fn2]{Author contribution statement is listed in Section \ref{sec:credit}.}
\fntext[fn2]{This research has been awarded the funding from the 2019 Computers \& Geosciences Research grant.}

\begin{abstract}
This paper presents a mixed finite element framework for coupled hydro-mechanical-chemical processes in heterogeneous porous media. 
The framework combines two types of locally conservative discretization schemes: 
(1) an enriched Galerkin method for reactive flow, 
and (2) a three-field mixed finite element method for coupled fluid flow and solid deformation.
This combination ensures local mass conservation, which is critical to flow and transport in heterogeneous porous media, with a relatively affordable computational cost.
A particular class of the framework is constructed for calcite precipitation/dissolution reactions, incorporating their nonlinear effects on the fluid viscosity and solid deformation. 
Linearization schemes and algorithms for solving the nonlinear algebraic system are also presented.
Through numerical examples of various complexity, we demonstrate that the proposed framework is a robust and efficient computational method for simulation of reactive flow and transport in deformable porous media,
even when the material properties are strongly heterogeneous and anisotropic.
\end{abstract}


\end{frontmatter}


\section{Introduction}

Hydro-mechanical-chemical (HMC) processes in porous media, in which fluid flow, solid deformation, and chemical reactions are tightly coupled, appear in a variety of problems ranging from groundwater and contaminant hydrology to subsurface energy production \cite{nick2013reactive,Hu2013,pandey2014investigation,pandey2017effect,nick2015mixed,choo2018cracking,tran2020coupling}. 
The multiphysical interactions in these problems give rise to strong heterogeneity in the material properties.
For instance, change in pore pressure perturbs effective stress in the solid matrix, which can, in turn, alter the conductivity and storability of the porous medium \cite{chen2007reservoir,Du2007,abou2013petroleum,kadeethum2019investigation,kadeethum2020well,nejati2019methodology}. 
Similarly, chemical processes can result in the precipitation or dissolution of solid minerals, which decreases or increases the pore volume, respectively, and thus, the conductivity \cite{salimzadeh2019effect,pandey2014investigation,pandey2017effect,rutqvist2017overview,ahkami2020lattice,choo2018cracking}. 
Therefore, accurate numerical modeling of coupled HMC problems requires a computational method that can robustly handle strong heterogeneity in porous media.

Numerical simulation of multiphysical problems in porous media has been a subject of extensive research (e.g. \cite{zhang2016mixed,dana2018convergence,kim2011stability,SlatlemVik2018,white2011block,nick2011hybrid,salinas2018discontinuous,chen2006computational,kadeethum2020finite,kadeethum2020pinn,Kadeethum2020ARMA}), 
and lots of software packages have been developed for this purpose. 
Notable examples include: (1) TOUGH software suite, which includes multi-dimensional numerical models for simulating the coupled thermo-hydro-mechanical-chemical (THMC) processes in porous and fractured media \cite{pruess1987tough,taron2009thermal,rutqvist2017overview,danko2012new}, 
(2) SIERRA Mechanics, which has simulation capabilities for coupling thermal, fluid, aerodynamics, solid mechanics and structural dynamics \cite{bean2012sierra}, 
(3) PyLith, a finite-element code for modeling dynamic and quasi-static simulations of coupled multiphysics processes \cite{aagaard2008pylith}, 
(4) OpenGeoSys project, which is developed mainly based on the finite element method using object-oriented programming THMC processes in porous media \cite{kolditz2012opengeosys}, 
(5) IC-FERST, a reservoir simulator based on control-volume finite element methods and dynamic unstructured mesh optimization \cite{adam2017dynamic,melnikova2016reservoir,obeysekara2016fluid,obeysekara2018modelling}, 
(6) DYNAFLOW\texttrademark, a nonlinear transient finite element analysis platform \cite{prevost1983dynaflow}, 
(7) DARSim, multiscale multiphysics finite volume based simulator \cite{ctene2016algebraic,cusini2015constrained,hosseinimehr2020adaptive}, 
(8) the CSMP, an object-oriented application program interface, for the simulation of complex geological processes, e.g.\ THMC, and their interactions  \cite{matthai2007numerical,salimzadeh2019coupled}, 
and 
(9) PorePy, an open-source modeling platform for multiphysics processes in fractured porous media \cite{keilegavlen2019porepy}.


Nevertheless, it remains challenging to simulate coupled HMC processes in porous media in a robust and efficient manner, especially when the material properties are highly heterogeneous and/or anisotropic.
Because HMC problems involve transport phenomena in heterogeneous porous media, the numerical method for these problems must ensure local (element-wise) conservation \cite{riviere2008discontinuous,lee2016locally}. 
The most practical method featuring local mass conservation may be the finite volume method with a standard two-point flux approximation scheme.
However, this standard finite volume method requires the grid to be aligned with the principal directions of the permeability/diffusivity tensors \cite{lipnikov2009local,choo2018cracking}, which inhibits the use of an unstructured grid when the permeability/diffusivity tensors are anisotropic. 
Multi-point flux-approximation methods have been developed to tackle this issue, but their implementation is often complicated and onerous \cite{choo2018large}. 
Discontinuous Galerkin (DG) methods offer an elegant way to handle arbitrarily anisotropic tensor conductivity/diffusivity. 
However, their computational cost is often impractical as a result of the proliferation of the degrees of freedom.

In this paper, we present a new framework for computational modeling of coupled HMC processes in porous media, which efficiently provides local mass conservation even when the material properties are strongly heterogeneous and anisotropic. 
The proposed framework combines two types of discretization methods:
(1) an enriched Galerkin (EG) method for reactive flow and transport, 
and (2) a three-field mixed finite element method for coupled hydro-mechanical processes. The EG method, which has recently been developed and advanced in the literature \cite{sun2009locally,lee2016locally,lee2018enriched,choo2018enriched,choo2018large,choo2019stabilized}, augments a piecewise constant function to the continuous Galerkin (CG) function space. 
This method uses the same interior penalty type form as the DG method, but it requires a substantially fewer number of degrees of freedom than the DG method. 
Thus the EG method can provide locally conservative solutions to the reactive flow system regardless of the grid--conductivity alignment.
For the hydro-mechanical sub-system of the HMC problem, we use a three-field mixed finite element formulation \cite{phillips2007coupling1, phillips2007coupling2, Haga2012}, which provides locally conservative, high-order solutions to the fluid velocity field.
Specifically, we employ the Lagrange finite elements for approximating the displacement field, the Brezzi-Douglas-Marini (BDM) element for the fluid velocity field, and the piecewise constant element for the fluid pressure field.
It is noted that this combination of elements is our personal choice, and one may use another combination for the same three primary fields as in \cite{ferronato2010fully,jha2007locally,Haga2012}. 

The purpose of this work is to develop an accurate numerical method for tackling coupled HMC processes in heterogeneous porous media, with a practically affordable computational cost.
Our specific objectives can be summarized as follows:
\begin{enumerate}
    \item To formulate a robust numerical approximation scheme for coupled HMC processes in heterogeneous porous media, employing a combination of locally conservative finite element methods.
    \item To reduce the computational cost for solving an advection-diffusion-reaction equation by using the EG method, which requires approximately two and three times fewer degrees of freedom than the DG method for 2D and 3D geometries, respectively \cite{KadNickLeeBallarin_2019_mixed}.
    \item To demonstrate the performance and capabilities of the proposed framework for modeling tightly coupled HMC problems with homogeneous to heterogeneous, isotropic to anisotropic permeability fields with local conservation.
\end{enumerate}



The rest of the paper is organized as follows. 
Section \ref{sec:methodology} describes the governing equations of coupled HMC processes. 
Section \ref{sec:numer} explains the discretization methods, linearization techniques, and solution algorithms of the proposed framework. 
Section \ref{sec:results} presents several numerical examples of various complexity and discusses key points found in this paper. 
Section \ref{sec:conclusion} concludes the work. \par

\section{Governing equations}\label{sec:methodology}

This section briefly describes all the equations used in this study, namely poroelasticity and advection-diffusion-reaction equations. 

Let $\Omega \subset \mathbb{R}^d$ ($d \in \{1,2,3\}$) denote the computational domain and $\partial \Omega$ denote the boundary. The time domain is denoted by $\mathbb{T} = \left(0,\mathrm{T}\right]$ with $\mathrm{T}>0$.
Primary variables used in this paper are $\bm{q} (\cdot , t) : \Omega \times  \mathbb{T} \to \mathbb{R}^d$, which is a vector-valued Darcy velocity (\si{m/s}), $p(\cdot ,  t) : \Omega \times  \mathbb{T} \to \mathbb{R}$, which is a scalar-valued fluid pressure (\si{Pa}), $\bm{u} (\cdot , t) : \Omega \times \mathbb{T} \to \mathbb{R}^d$, which is a vector-valued displacement (\si{m}), $c_i : \Omega \times \mathbb{T} \rightarrow \mathbb{R}$, which is the $i$-th component of chemical concentration (\si{m mol/m^3}).

\subsection{Poroelasticity}
To begin, we adopt Biot's poroelasticity theory for coupled hydro-mechanical processes in porous media \cite{biot1941general,biot1957elastic}.
Although poroelasticity may oversimplify deformations in soft porous materials such as soils \cite{Choo2016,Borja2016,Macminn2016,Zhao2020}, it would be reasonably good for stiff materials such as rocks, which is the focus of this work. 
The poroelasticity theory provides two coupled governing equations, namely linear momentum and mass balance equations.
Under quasi-static conditions, the linear momentum balance equation can be written as
\begin{equation}
\nabla \cdot \bm{\sigma} (\bm{u},p) + \bm{f} = \bm{0},
\end{equation}
where $\bm{f}$ is the body force term defined as $\rho \phi \mathbf{g}+\rho_{s}(1-\phi) \mathbf{g}$, where $\rho$ is the fluid density, $\rho_s$ is the solid density, $\phi$ is the porosity, $\mathbf{g}$ is the gravitational acceleration vector. 
The gravitational force will be neglected in this study, but the body force term will be kept in the succeeding formulations for a more general case.
Further, $\bm{\sigma}$ is the total stress tensor, which may be related to the effective stress tensor $\bm{\sigma}^{\prime}$ and the pore pressure $p$ as
\begin{equation}
\bm{\sigma} (\bm{u},p) = \bm{\sigma}^{\prime}(\bm{u}) - \alpha p \mathbf{I}.
\end{equation}
Here, $\mathbf{I}$ is the second-order identity tensor, and 
$\alpha$ is the Biot coefficient defined as \cite{jaeger2009fundamentals}:
\begin{equation} \label{eq:biot_coeff}
\alpha = 1-\frac{K}{K_{{s}}},
\end{equation}
\noindent
with {$K$} and {$K_s$} being the bulk moduli of the solid matrix and the solid grain, respectively. 
According to linear elasticity, the effective stress tensor has a constitutive relationship with the displacement vector, which can be written as
\begin{equation}
\bm{\sigma}^{\prime}(\bm{u}) =
\lambda_{l}  \tr(\bm{\varepsilon}(\bm{u})) \mathbf{I}
+ 2 \mu_{l} \bm{\varepsilon}{(\bm{u})}.
\end{equation} 
\noindent
Here, $\bm{\varepsilon}$ is the infinitesimal strain tensor, defined as
\begin{equation}
\bm{\varepsilon}(\bm{u}) :=\frac{1}{2}\left(\nabla \bm{u}+(\nabla \bm{u})^{\intercal}\right),
\end{equation}
and $\lambda_{l}$ and $\mu_{l}$ are the Lam\'e constants, which are related to the bulk modulus and the Poisson ratio $\nu$ of the solid matrix as

\begin{equation}\label{eq:lambda_l}
\lambda_{l}=\frac{3 K \nu}{1+\nu}, \text{ and } \mu_{l}=\frac{3 K(1-2 \nu)}{2(1+\nu)}.
\end{equation}
\noindent

For this solid deformation problem, the domain boundary $\partial \Omega$ is assumed to be suitably decomposed into displacement and traction boundaries, $\partial \Omega_u$ and $\partial \Omega_{t}$, respectively. 
Then the linear momentum balance equation is supplemented by the boundary and initial conditions as:

\begin{equation} \label{eq:linear_balance}
\begin{split}
\nabla \cdot \bm{\sigma}^{\prime}(\bm{u}) +\alpha \nabla \cdot \left(p \mathbf{I}\right) 
+ \bm{f} = \bm{0}  &\text { \: in \: } \Omega \times \mathbb{T}, \\
\bm{u} =\bm{u}_{D} &\text { \: on \: } \partial \Omega_{u} \times \mathbb{T},\\
\bm{\sigma} {(\bm{u})} \cdot  \mathbf{n}=\bm{t}_{D} &\text { \: on \: } \partial \Omega_{t} \times \mathbb{T}, \\
\bm{u}=\bm{u}_{0}  &\text { \: in \: } \Omega \text { at } t = 0,
\end{split}
\end{equation}

\noindent
where $\bm{u}_D$ and ${\bm{t}_D}$ are prescribed displacement and traction values at the boundaries, respectively, and $\mathbf{n}$ is the unit normal vector to the boundary.

Next, the mass balance equation is given as \cite{coussy2004poromechanics,kim2011stability,pandey2017effect,salimzadeh2019coupled}:
\begin{equation} \label{eq:mass_balance_old}
\frac{1}{M} \dfrac{\partial p}{\partial t} + 
\alpha  \frac{\partial {\varepsilon_{v}}}{\partial t}  + \dfrac{\partial \phi_c}{\partial t}
+ \nabla \cdot  \bm{q} = g \text { in } \Omega \times \mathbb{T}, 
\end{equation}

\noindent
where
\begin{equation} \label{eq:1/M}
\frac{1}{M}  = \left(\phi_0 c_{f}+\dfrac{\alpha-\phi_0}{K_{s}}\right)
\end{equation}
is the Biot modulus.
\noindent
Here, $c_f$ is the fluid compressibility, $\phi_0$ is the initial porosity, ${\varepsilon_{v}}$ := $\operatorname{tr}(\bm{\varepsilon}) = \nabla \cdot \bm{u}$ is the volumetric strain, and $g$ is a sink/source term. 
Because we will introduce chemical effects later on, we have added $\frac{\partial \phi_c}{\partial t}$ to the standard poroelasticity equation \cite{chaudhuri2013early,pandey2014investigation,pandey2017effect,salimzadeh2019coupled}.
This term will be discussed again after introducing chemical effects. 
Also, $\bm{q}$ is the superficial velocity vector, which is given by Darcy's law as
\begin{equation} \label{eq:darcy}
\bm{q} =- \frac{\bm{k}(\phi)}{\mu(c_i)} (\nabla p-\rho \mathbf{g}).
\end{equation}
\noindent
Note that here the fluid viscosity $\mu$ is considered a function of concentration $c_i$. 
Again, the gravitational force, $\rho \mathbf{g}$, will be neglected in this work, without loss of generality. 
In addition, $\bm{k}(\phi)$ is the matrix permeability tensor defined as

\begin{equation} \label{eq:permeability_matrix}
\bm{k} := 
\begin{cases}
k_{mult}(\phi) \left[ \begin{array}{lll}{{k}^{xx}} & {{k}^{xy}} & {{k}^{xz}} \\ {{k}^{yx}} & {{k}^{yy}} & {{k}^{yz}} \\ {{k}^{zx}} & {{k}^{zy}} & {k}^{zz}\end{array}\right] & \text{if} \ d = 3, \ \\ \\
k_{mult}(\phi) \left[ \begin{array}{ll}{{k}^{xx}} & {{k}^{xy}}  \\ {{k}^{yx}} & {{k}^{yy}} \\ \end{array}\right]  & \text{if} \ d = 2, \ \\ \\
k_{mult}(\phi) \ k  & \text{if} \ d = 1,
\end{cases}
\end{equation}


\noindent
The $k^{xx}$, $k^{yy}$, and $k^{zz}$ represent the matrix permeability in $x$-, $y$-, and $z$-direction, respectively. The $k_{mult}(\phi)$ is a multiplier used to update $\bm{k}$ when $\phi$ is altered, which will be described later.

For the fluid flow problem, the domain boundary $\partial \Omega$ is also suitably decomposed into the pressure and flux boundaries,  $\partial \Omega_p$ and $\partial \Omega_q$, respectively.
In what follows, we apply the fixed stress split scheme \cite{kim2011stability,mikelic2013convergence}, assuming $\left(\sigma_{v}-\sigma_{v, 0}\right)+\alpha \left(p-p_{0}\right)=K \varepsilon_{v}$. 
Then we write the fluid flow problem with boundary and initial conditions as


\begin{equation} \label{eq:mass_balance}
\begin{split}
\left(\frac{1}{M}+\frac{\alpha^{2}}{K}\right) \frac{\partial p}{\partial t}+\frac{\alpha}{K} \frac{\partial \sigma_{v}}{\partial t}+ \dfrac{\partial \phi_c}{\partial t}+\nabla \cdot \bm{q} = g  &\text { \: in \: } \Omega \times \mathbb{T}, \\
p=p_{D} &\text { \: on \: } \partial \Omega_{p} \times \mathbb{T}, \\
\bm{q} \cdot \mathbf{n}=q_{D} &\text { \: on \:} \partial \Omega_{q} \times \mathbb{T}, \\
p=p_{0} &\text { \: in \: } \Omega \text { at } t = 0,
\end{split}
\end{equation}

\noindent
where $\sigma_{v}:=\frac{1}{3} \tr(\bm{\sigma})$ is the volumetric stress, and $p_D$ and $q_D$ are the given boundary pressure and flux, respectively.

\subsection{Reactive flow}

An advection-diffusion-reaction system for $N_c$ number of the miscible species is given by the following equations. For all $i=1,\hdots,N_c$,
\begin{equation}
\frac{\partial}{\partial t}(\phi c_i) + \nabla \cdot \eta(\bm{q}, c_i) = q_i(c_i),\ \mbox{ in } \Omega \times {\mathbb{T}},
\label{eqn:main_transport}
\end{equation}

\noindent
where $q_i(c_i)$ is a reaction term coupled with sink/source for each component,
and the mass flux $\eta(\bm{q},c_i)$ is defined as
\begin{equation}
\eta(\bm{q},c_i) :=  \bm{q} c_i - {\bm{D}_{e,i}}(\phi) \nabla c_i.
\end{equation}

\noindent
Here $ {\bm{D}_{e,i}}(\phi)$ is the effective diffusion coefficient tensor defined as 
\begin{equation}\label{eq:D_update}
{\bm{D}_{e,i}} := \frac{\phi}{\tau} \bm{D}_{i},
\end{equation}

\noindent
where $\tau=\phi^{-\frac{1}{2}}$ \cite{tjaden2016origin,mu2008determination} and $\bm{D}_{i}$ is the given diffusion coefficient tensor. 
The boundary for the advection-diffusion-reaction system is decomposed into inflow and outflow boundaries, denoted by $\partial \Omega_{\rm in}$ and $\partial \Omega_{\rm out}$, respectively, which are defined as

\begin{equation}
\partial \Omega_{\rm in} := \{ \bm{x} \in \partial \Omega : \bm{q} \cdot \mathbf{n} < 0\} \quad \mbox{ and } \quad \partial \Omega_{\rm out} := \{ \bm{x} \in \partial \Omega : \bm{q} \cdot \mathbf{n} \geq 0\}.
\label{eq:in_and_out}
\end{equation}

In what follows, we specialize the model to calcite precipitation and dissolution reactions, which requires us to solve a calcite-carbonic acid system.  
In general, the system requires eight transport equations to solve the concentration values of the following  main species/ions: $\left\{\mathrm{H}^{+}\right.$, $\mathrm{Ca}^{2+}$, $\mathrm{CaHCO}_{3}^{+}$, $\mathrm{OH}^{-}$, $\mathrm{CO}_{3}^{2-}$, $\mathrm{HCO}_{3}^{-}$, $\mathrm{H}_{2} \mathrm{CO}_{3}^{*}$, $\mathrm{CaCO}_{3}^{*}(\mathrm{Aq})\}$
\cite{chaudhuri2013early,pandey2014investigation,raoof2013poreflow,morel1993principles}. 
For simplicity, in this paper we consider a reduced system 
based on the empirical relationship presented in \cite{chaudhuri2013early,pandey2014investigation,pandey2017effect}, in which $N_c$ decreases to 1. 
Thus, letting $c := c_1$, we write the advection-diffusion-reaction system with its boundary and initial conditions as follows:


\begin{equation} \label{eq:transport}
\begin{split}
\frac{\partial}{\partial t}(\phi c) + \nabla \cdot \left( \bm{q} c  \right)- \nabla \cdot \left( \bm{D}_{e}(\phi) \nabla c \right) = q  &\mbox{ \: in \: } \Omega \times (0,\mathbb{T}], \\
\eta(\bm{q},c) \cdot \mathbf{n} = {c_{in}} \bm{q}\cdot \mathbf{n} &\mbox{ \: on \: } \partial \Omega_{{\rm in}} \times (0,\mathbb{T}],  \\
{\bm{D}_{e}}(\phi) \nabla c \cdot \mathbf{n} = 0  &\mbox{ \: on \: } \partial \Omega_{{\rm out}} \times (0,\mathbb{T}], \\
c={c_{0}}  &\text{ \: in \: } \Omega \text { at } t = 0,
\end{split}
\end{equation}

\noindent
where $c_{in}$ is the inflow concentration, $c_{0}$ is the initial concentration, and $q$ represents a source term reflecting the calcite dissolution/precipitation reactions.
For this term, here we adopt the term in \cite{chaudhuri2013early,pandey2014investigation,pandey2017effect}, given by
\begin{equation}
q=R_{c} A_{s},
\end{equation}
\noindent
where $A_s$ is the specific surface of the porous medium, and $R_{c}$ is the reaction rate calculated as
\begin{equation} R_c=\left\{\begin{array}{cl}
10^{r}, & \text { for } \widetilde{c}>0, \\
-10^{r}, & \text { for } \widetilde{c}<0, \\
0.0, & \text { for } \widetilde{c}=0,
\end{array}\right.\end{equation}

\noindent
with

\begin{equation}\label{eq:c_tilde}
\widetilde{c}=\frac{\left(c_{e q}-c\right)}{c_{e q}},
\end{equation}

\begin{equation}\begin{aligned}\label{eq:small_r}
r=a_{0}+a_{1} \tau+a_{2} \log |\widetilde{c}|+a_{3} \tau^{2}+a_{4} \tau \log |\widetilde{c}|+a_{5}(\log |\widetilde{c}|)^{2},
\end{aligned}\end{equation}

\noindent
and

\begin{equation}\begin{aligned}\label{eq:c_sat}
c_{e q} &=1.417 \times 10^{-3}+3.823 \times 10^{-6} p-4.313 \times 10^{-7} \tau \\
&-2.148 \times 10^{-8} p^{2}+4.304 \times 10^{-8} p \tau-7.117 \times 10^{-8} \tau^{2}.
\end{aligned}\end{equation}

\noindent
Here, $\tau$ is the medium temperature, and $a_{0}$, $a_{1}$, $a_{2}$, $a_{3}$, $a_{4}$, and $a_{5}$ are defined in Table \ref{tab:calcite_coeff}. 

\begin{table}[!ht]
  \centering
  \caption{Coefficients of the \eqref{eq:small_r} for different range of \eqref{eq:c_tilde}}
    \begin{tabular}{|l|c|c|c|c|c|c|}
    \hline & $a_{0}$ & $a_{1}$ & $a_{2}$ & $a_{3}$ & $a_{4}$ & $a_{5}$ \\
\hline $\widetilde{c}>0.01$ & -5.73 & $1.25 \times 10^{-2}$ & 1.38 & $2.61 \times 10^{-5}$ & $-4.01 \times 10^{-3}$ & $3.26 \times 10^{-1}$ \\
$-0.01<\widetilde{c}$ & -6.45 & $2.09 \times 10^{-2}$ & $-4.65 \times 10^{-2}$ & $3.06 \times 10^{-5}$ & $9.25 \times 10^{-3}$ & $-4.59 \times 10^{-1}$ \\
$-0.01<\widetilde{c} \leq 0.01$ & -5.80 & $1.35 \times 10^{-2}$ & $9.97 \times 10^{-1}$ & $3.80 \times 10^{-5}$ & $1.51 \times 10^{-5}$ & $-4.87 \times 10^{-4}$ \\
    \hline
    \end{tabular}%
  \label{tab:calcite_coeff}%
\end{table}%

Before closing this section, we describe physical properties that are coupled with primary variables, $\bm{u}$, $\bm{q}$, $p$, and $c$. 
The porosity change due to solid deformation may be expressed as \cite{biot1941general,dana2018multiscale,dana2018convergence}:

\begin{equation}\label{eq:porosity_by_mech}
\begin{split}
\phi_m=\phi_{0}+\left(\alpha-\phi_{0}\right)\left(\epsilon_{v}-\epsilon_{v_{0}}\right)+\frac{\left(\alpha-\phi_{0}\right)(1-\alpha)}{K}\left(p-p_{0}\right),
\end{split}
\end{equation}

\noindent
where $\epsilon_{v_{0}}$ is the initial volumetric strain. 
The porosity alteration due to calcite dissolution/precipitation is calculated as
\begin{equation}\label{eq:porosity_by_chem}
\Gamma \left( \bm{u}, c \right) = \frac{\partial \phi_c}{\partial t}=\frac{R_c A_{s}}{\rho_{s} \omega},
\end{equation}

\noindent
where $\omega$ is the number of moles of total precipitated species per kilogram of rock (assumed to be 10.0 in this study following \cite{pandey2014investigation,pandey2017effect}), and $\rho_{s}=2500$ \si{kg/m^3} is used for throughout this paper. 
Note that this term, \eqref{eq:porosity_by_chem}, enters \eqref{eq:mass_balance}. 
Also, the terms $\phi_m$ and $\phi_c$ are used to distinguish between the changes in $\phi$ due to solid deformation as in \eqref{eq:porosity_by_mech}, and chemical reactions as in \eqref{eq:porosity_by_chem}, respectively. 
The changes in porosity due to \eqref{eq:porosity_by_mech} and \eqref{eq:porosity_by_chem}, also affect the specific surface for porous medium ($A_s$) as
\begin{equation}\label{eq:a_s_update}
A_{s}=A_{0} \frac{\phi}{\phi_{0}} \frac{\log (\phi)}{\log \left(\phi_{0}\right)},
\end{equation}

\noindent
where $A_{0}$ is the initial value of $A_s$, and it is set as 5000 throughout this study \cite{taheriotaghsara2020prediction}. 
Furthermore, the porosity change influences the matrix permeability as \cite{rutqvist2002modeling,rutqvist2003role,min2004stress}:

\begin{equation}\label{eq:perm_update_rut}
\bm{k}= \bm{k}_0 k_{mult}(\phi) =  \bm{k}_0 \exp \left(b \left( \frac{\phi}{\phi_{0}} -1\right)\right),
\end{equation}

\noindent
where $\bm{k}_0$ is the initial matrix permeability and $b$ is an empirical parameter determined experimentally.
In this work, we set $b = 22.2$ following \cite{rutqvist2002modeling}. 
The change in $c$ also affects $\mu$, 
and we adopt the specific form from \cite{grolimund2001aggregation,bijeljic2007pore,yortsos2006selection}, given by

\begin{equation}\label{eq:mu_update}
\mu= \log \left(\mu_{l}\right) + \left( \frac{c-c_l}{c_h-c_l} \right) \left( \log \left(\mu_h \right) - \log \left( \mu_l \right) \right),
\end{equation}

\noindent
where $c_l$ and $c_h$ are lower and higher bounds of the concentration, and $\mu_l$ and $\mu_h$ are fluid viscosity corresponding to $c_l$ and $c_h$, respectively. 
Table \ref{tab:sum_effect} summarizes the effects of physical processes on material properties considered in this study. 
Note that the numbers, e.g., \eqref{eq:porosity_by_mech}, point out the equations used to represent these effects, while a hyphen means the absence of a relationship.

\begin{table}[!ht]
\centering
  \caption{Summary of the effects of individual physical processes on physical properties}
\begin{tabular}{|l|c|c|c|}
\hline
              Physical properties                  & Mechanical deformation & Fluid pressure & Calcite concentration \\ \hline
$\phi$                        &  \eqref{eq:porosity_by_mech}                       & \eqref{eq:porosity_by_mech}                 &  \eqref{eq:porosity_by_chem}                      \\ \hline
$\bm{k}$                    &  \eqref{eq:porosity_by_mech} + \eqref{eq:perm_update_rut}                       & \eqref{eq:porosity_by_mech} + \eqref{eq:perm_update_rut}                 &  \eqref{eq:porosity_by_chem} + \eqref{eq:perm_update_rut}                      \\ \hline
$\mu$                 & -                      & -              &  \eqref{eq:mu_update}                    \\ \hline
$\bm{D}_{e}$ &   \eqref{eq:porosity_by_mech} + \eqref{eq:D_update}                   & \eqref{eq:porosity_by_mech} + \eqref{eq:D_update}                 &  \eqref{eq:porosity_by_chem} + \eqref{eq:D_update}                      \\ \hline
$A_s$                &   \eqref{eq:porosity_by_mech} +  \eqref{eq:a_s_update}                        & \eqref{eq:porosity_by_mech} +  \eqref{eq:a_s_update}              &     \eqref{eq:porosity_by_chem}   + \eqref{eq:a_s_update}                     \\ \hline
\end{tabular}
\label{tab:sum_effect}%
\end{table}

\section{Numerical methods}\label{sec:numer}

In this section, we describe the numerical methods for the governing system described in the previous sections. Here, we utilize a combination of a mixed finite element method for spatial discretization, and employ both a backward differentiation formula and an explicit Runge-Kutta method for temporal discretization.

\subsection{Domain discretization and geometrical quantities} 

We begin by introducing the notations used throughout this paper. 
Let $\mathcal{T}_h$ be a shape-regular triangulation obtained by a partition of $\Omega$ into $d$-simplices (triangles in $d=2$, tetrahedra in $d=3$). For each cell $T \in \mathcal{T}_h$, we denote by $h_{T}$ the diameter of $T$, and we set $h=\max_{T \in \mathcal{T}_h} h_{T}$ and $h_{l}=\min_{T \in \mathcal{T}_h} h_{T}$.
We further denote by $\mathcal{E}_h$ the set of all faces (i.e., $d - 1$ dimensional entities connected to at least a $T \in \mathcal{T}_h$) and by $\mathcal{E}_h^{I}$ and $\mathcal{E}_h^{\partial}$ the collection of all interior and boundary facets, respectively. 
The boundary set $\mathcal{E}_h^{\partial}$ is decomposed into two disjoint subsets associated with the Dirichlet boundary faces, and the Neumann boundary faces for each of \eqref{eq:linear_balance} and \eqref{eq:mass_balance}. 
In particular, $\mathcal{E}_{h}^{D,u}$ and $\mathcal{E}_{h}^{N,u}$ correspond to the faces on $\partial \Omega_u$ and $\partial \Omega_{tr}$, respectively,  for \eqref{eq:linear_balance}. 
On the other hand, for \eqref{eq:mass_balance},  $\mathcal{E}_{h}^{D,m}$ and $\mathcal{E}_{h}^{N,m}$ conform to $\partial \Omega_p$ and $\partial \Omega_{q}$, respectively. Lastly, for  \eqref{eq:transport}, $\mathcal{E}_h^{\partial}$ is decomposed into $\mathcal{E}_{h}^{\mathrm{In}}$ and $\mathcal{E}_{h}^{\mathrm{Out}}$.

We also define
$$
e = \partial T^{+}\cap \partial T^{-}, \ \ e \in \mathcal{E}_h^I,
$$
\noindent
where  $T^{+}$ and $T^{-}$ are the two neighboring elements to $e$. We denote by $h_e$ the characteristic length of $e$ calculated as
\begin{equation}
h_{e} :=\frac{\operatorname{meas}\left(T^{+}\right)+\operatorname{meas}\left(T^{-}\right)}{2 \operatorname{meas}(e)},
\end{equation}

\noindent
depending on the argument, meas($\cdot$) represents the measure of a cell or of a facet.

Let $\mathbf{n}^{+}$ and $\mathbf{n}^{-}$ be the outward unit normal vectors to  $\partial T^+$ and $\partial T^-$, respectively. 
For any given scalar function $\zeta: \mathcal{T}_h \to \mathbb{R}$ and vector function $\bm{\tau}: \mathcal{T}_h \to \mathbb{R}^d$, we denote by $\zeta^{\pm}$ and $\bm{\tau}^{\pm}$ the restrictions of $\zeta$ and $\bm{\tau}$ to $T^\pm$, respectively. 
Subsequently, we define the weighted average operator as

\begin{equation}
\{\zeta\}_{\delta e}=\delta_{e} \zeta^{+}+\left(1-\delta_{e}\right) \zeta^{-}, \ \text{ on } e \in \mathcal{E}_h^I, 
\end{equation}

\noindent
and

\begin{equation}
\{\bm{\tau}\}_{\delta e}=\delta_{e} \bm{\tau}^{+}+\left(1-\delta_{e}\right) \bm{\tau}^{-},
\ \text{ on } e \in \mathcal{E}_h^I,
\end{equation}

\noindent
where $\delta_{e}$ is calculated by \cite{ErnA_StephansenA_ZuninoP-2009aa,ern2008posteriori}:

\begin{equation}
\delta_{e} :=\frac{{k}^{-}_e}{{k}^{+}_e+{k}^{-}_e}.
\end{equation}
Here, 
\begin{equation}
{k}^{+}_e :=\left(\mathbf{n}^{+}\right)^{\intercal} \cdot \bm{k}^{+}  \mathbf{n}^{+}, \ \text{ and }
{k}^{-}_e :=\left(\mathbf{n}^{-}\right)^{\intercal} \cdot \bm{k}^{-}  \mathbf{n}^{-},
\end{equation}
where  ${k_e}$ is a harmonic average of $k^{+}_e$ and ${k}^{-}_e$ which reads
\begin{equation}
{k_{e}}:= \frac{2{k}^{+}_e {k}^{-}_e}{{k}^{+}_e+{k}^{-}_e},
\end{equation}
and $\bm{k}$ is defined as in \eqref{eq:permeability_matrix}.
\noindent
The jump across an interior edge will be defined as 
\begin{align*}
\jump{\zeta} = \zeta^+\mathbf{n}^++\zeta^-\mathbf{n}^- \quad \mbox{ and } \quad \jtau = \bm{\tau}^+\cdot\mathbf{n}^+ + \bm{\tau}^-\cdot\mathbf{n}^- \quad \mbox{on } e\in \mathcal{E}_h^I. 
\end{align*}

Finally, for $e \in \mathcal{E}^{\partial}_h$, we set $\av{\zeta}_{\delta_e} :=   \zeta$ and $\av{\bm{\tau}}_{\delta_e} :=  \bm{\tau}$ for what concerns the definition of the weighted average operator, and $\jump{\zeta} :=  \zeta \mathbf{n}$ and $\jump{\bm{\tau}} :=  \bm{\tau} \cdot \mathbf{n}$ as definition of the jump operator.

\subsection{Temporal discretization}

The time domain $\mathbb{T} = \left(0, \mathrm{T}\right]$ is partitioned into $N$ subintervals such that $0=: t^{0}<t^{1}<\cdots<t^{N} := \mathrm{T}$. The length of each subinterval $\Delta t^{n-1}$ is defined as $\Delta t^{n-1}=t^{n}-t^{n-1}$ where $n$ represents the current time step. We assume that the user provides the initial $\Delta t^0$, while an adaptive procedure is carried out to choose $\Delta t^{n-1}$, $n > 1$, as follows:

\begin{equation} \label{eq:time_mult}
\Delta t^{n-1} :=
\begin{cases}
\mathrm{CFL} \frac{h_l}{\left\|\bm{q}^{n-1}\right\|_{\infty}} & \text{if} \ \Delta t^n \le \Delta t_{\max} \ \\
 \Delta t_{\max}  & \text{if} \ \Delta t^n > \Delta t_{\max},
\end{cases}
\end{equation}

\noindent
where $\mathrm{CFL}$ is a constant that the user can provide according to the Courant-Friedrichs-Lewy condition \cite{courant1967partial}, $\left\|\cdot\right\|_{\infty}$ is the maximum norm of a vector function, and $\Delta t_{\max}$ is a maximum allowed time step.
Note that we use $\Delta t_{\max}$ as a tool to control $\Delta t^n$ as the model approaches a steady-state condition since $\left\|\bm{q}^{n-1}\right\|_{\infty}$ may approach zero, which would lead to a very large ratio $\frac{h_l}{\left\|\bm{q}^{n-1}\right\|_{\infty}}$. 

Let $\varphi(\cdot, t)$ be a scalar function and $\varphi^{n}$ be its approximation at time $t^n$, i.e. $\varphi^{n} \approx \varphi\left(t^{n}\right)$. We employ the following backward differentiation formula \cite{ibrahim2007implicit,akinfenwa2013continuous,lee2018phase}

\begin{equation} \label{eq:bdf_gen}
\mathrm{BDF}_{m}\left(\varphi^{n}\right):=\left\{\begin{array}{ll}
\frac{1}{\Delta t^n}\left(\varphi^{n}-\varphi^{n-1}\right) & m=1 \\
\frac{1}{2\Delta t^n}\left(3 \varphi^{n}-4 \varphi^{n-1}+\varphi^{n-2}\right) & m=2 \\
\frac{1}{6\Delta t^n}\left(11 \varphi^{n} -18 \varphi^{n-1}+9 \varphi^{n-2}-2\varphi^{n-3}\right) & m=3 \\
\frac{1}{12\Delta t^n}\left(25 \varphi^{n} -48 \varphi^{n-1}+36 \varphi^{n-2}-16\varphi^{n-3}+3\varphi^{n-4}\right) & m=4
\end{array}\right.
\end{equation}
for the discretization of the time derivative of $\varphi(\cdot, t)$ at time $t^n$. We also utilize the explicit Runge-Kutta methods \cite{dormand1980family,chen2006computational}:

\begin{equation}
\mathrm{RK}_1(\varphi^{n}) = \varphi^{n+1}=\varphi^{n}+\kappa_{1},
\end{equation}

\begin{equation*}\begin{array}{l}
\kappa_{1}=\Delta t^n F\left(\mathbb{X}^{n}, \mathbb{Y}^{n}\right),
\end{array}\end{equation*}

\noindent
for the first order Runge-Kutta method corresponding to the explicit Euler method, and

\begin{equation}
\mathrm{RK}_4(\varphi^{n}) = \varphi^{n+1}=\varphi^{n}+\frac{1}{6} \kappa_{1}+\frac{1}{3} \kappa_{2}+\frac{1}{3} \kappa_{3}+\frac{1}{6} \kappa_{4},
\end{equation}

\begin{equation*}\begin{array}{l}
\kappa_{1}=\Delta t^n F\left(\mathbb{X}^{n}, \mathbb{Y}^{n}\right), \\
\kappa_{2}=\Delta t^n F\left(\mathbb{X}^{n}+\frac{1}{2} \Delta t^n, \mathbb{Y}^{n}+\frac{1}{2} \kappa_{1}\right), \\
\kappa_{3}=\Delta t^n F\left(\mathbb{X}^{n}+\frac{1}{2} \Delta t^n, \mathbb{Y}^{n}+\frac{1}{2} \kappa_{2}\right), \\
\kappa_{4}=\Delta t^n F\left(\mathbb{X}^{n}+\Delta t^n, \mathbb{Y}^{n}+\kappa_{3}\right),
\end{array}\end{equation*}

\noindent
for the forth order Runge-Kutta method, $F\left(\mathbb{X}^{n}, \mathbb{Y}^{n}\right)$ is any functions with independent variable $\mathbb{X}$ and dependent variable $\mathbb{Y}$ \cite{dormand1980family,chen2007reservoir}, which we will specify in the linearization and solving processes in Section \ref{sec:split}. 

Finally, we define an extrapolation operator as follows \cite{chen2006computational,chen2007reservoir}:

\begin{equation} \label{eq:gen_extrapolate}
\mathrm{EX}\left( \varphi \right) = \hat{\varphi}^{n+1} =
\begin{cases}
\left(1+\dfrac{\Delta t^{n}}{\Delta t^{n-1}}\right) {\varphi}^{n}-\dfrac{\Delta t^{n}}{\Delta t^{n-1}} {\varphi}^{n-1} & \text{if} \ n \geq 1, \ \\
 {\varphi}^{0}  & \text{if} \ n=0,
\end{cases}
\end{equation}

\noindent
and in the following we will adopt the notation $\hat{\varphi}^{n+1}$ to denote an extrapolation value of $\{\varphi^{n}, \varphi^{n-1}\}$.

\subsection{Spatial discretization}\label{sec:space_discretize}

In this framework, the displacement field is approximated by the classical continuous Galerkin method (CG) method, 
and the fluid velocity and pressure fields are discretized by 
the Brezzi-Douglas-Marini (BDM) element \cite{brezzi2012mixed} 
and the piecewise constants discontinuous Galerkin (DG) method, respectively, to ensure local mass conservation. 
Lastly, the concentration field is discretized by the enriched Galerkin (EG) method \cite{lee2016locally,sun2009locally}.


To begin, we define the finite element space for the CG function space for a vector-valued function:
\begin{equation}
 \mathcal{U}_{h}^{\mathrm{CG}_{k}}\left(\mathcal{T}_{h}\right) :=\left\{\bm{\psi_u} \in \mathbb{C}^{0}(\Omega{; \mathbb{R}^d}) :\left.\bm{\psi_u}\right|_{T} \in \mathbb{P}_{k}(T{; \mathbb{R}^d}), \forall T \in \mathcal{T}_{h}\right\},
\label{eq:CG_U}
\end{equation}
\noindent
where $\mathbb{C}^0(\Omega{; \mathbb{R}^d})$ denotes the space of vector-valued piecewise continuous polynomials, $\mathbb{P}_{k}(T{; \mathbb{R}^d})$ is the space of polynomials of degree at most $k$ over each element $T$, and $\bm{\psi_u}$ denotes a generic function of $\mathcal{U}_{h}^{\mathrm{CG}_{k}}\left(\mathcal{T}_{h}\right)$. 
In addition, the CG space for scalar-valued functions is defined as:
\begin{equation}
\mathcal{P}_{h}^{\mathrm{CG}_{k}}\left(\mathcal{T}_{h}\right) :=\left\{\psi_p \in \mathbb{C}^{0}(\Omega) :\left.\psi_p\right|_{T} \in \mathbb{P}_{k}(T), \forall T \in \mathcal{T}_{h}\right\},
\label{eq:CG_space_s}
\end{equation}

\noindent
where $\mathbb{C}^{0}(\Omega) := \mathbb{C}^{0}(\Omega; \mathbb{R})$ and $\mathbb{P}_{k}(T) := \mathbb{P}_{k}(T{; \mathbb{R}})$.
Next, we define the following DG function space:

\begin{equation}\label{eq:DG_P}
\mathcal{P}_{h}^{\mathrm{DG}_{k}}\left(\mathcal{T}_{h}\right) :=\left\{\psi_p \in L^{2}(\Omega) :\left.\psi_p\right|_{T} \in \mathbb{P}_{k}(T), \forall T \in \mathcal{T}_{h}\right\},
\end{equation}

\noindent
where $L^{2}(\Omega)$ is the space of square-integrable scalar functions. This non-conforming finite element space allows us to consider discontinuous functions and coefficients rigorously. We then define the EG finite element space with polynomial order $k$ as:
\begin{equation}\label{eq:EG_P}
\mathcal{P}_{h}^{\mathrm{EG}_{k}}\left(\mathcal{T}_{h}\right) :=\mathcal{P}_{h}^{\mathrm{CG}_{k}}\left(\mathcal{T}_{h}\right) \oplus \mathcal{P}_{h}^{\mathrm{DG}_{0}}\left(\mathcal{T}_{h}\right),
\end{equation}

\noindent
i.e., a CG finite element space enriched by the space $\mathcal{P}_{h}^{\mathrm{DG}_{0}}\left(\mathcal{T}_{h}\right)$ of piecewise constant functions. In the following we denote ${\psi_c}$ a generic function of $\mathcal{P}_{h}^{\mathrm{EG}_{k}}\left(\mathcal{T}_{h}\right)$.

Lastly, we define the BDM function space as follows \cite{brezzi2012mixed}:

\begin{equation}
\mathcal{V}_{h}^{\mathrm{BDM}_{k}}\left(\mathcal{T}_{h}\right) :=\left\{\bm{\psi_v} \in H(\operatorname{div}, \Omega):\left.\bm{\psi_v}\right|_{T} \in \mathrm{BDM}(T), \forall T \in \mathcal{T}_{h}\right\} 
\label{eq:BDM_V}
\end{equation}
\noindent
where $\bm{\psi_v}$ denotes a generic function of $\mathcal{V}_{h}^{\mathrm{BDM}_{k}}\left(\mathcal{T}_{h}\right)$ and $\mathrm{BDM}(T)$ is defined according to \cite{brezzi2012mixed}.

\subsubsection{Fully discrete form}
We now present the fully discrete form of the coupled HMC problem using the above-described combination of finite element spaces.
%
In particular, we seek the approximated displacement solution $\bm{u_h} \in \mathcal{U}_{h}^{\mathrm{CG}_{2}}\left(\mathcal{T}_{h}\right)$
as done in \cite{choo2018enriched,Kadeethum2019ARMA,SlatlemVik2018},
fluid pressure $p_h \in \mathcal{P}_{h}^{\mathrm{DG}_{0}}\left(\mathcal{T}_{h}\right)$,
velocity approximation $\bm{q}_h \in \mathcal{V}_{h}^{\mathrm{BDM}_{1}}\left(\mathcal{T}_{h}\right)$,
and concentration approximation $c_h \in \mathcal{P}_{h}^{\mathrm{EG}_{1}}\left(\mathcal{T}_{h}\right)$.


We multiply the linear momentum balance equation \eqref{eq:linear_balance}
by a test function $\bm{\psi_u} \in \mathcal{U}_{h}^{\mathrm{CG}_{2}}\left(\mathcal{T}_{h}\right)$.
The fully discretized linear momentum balance equation thus has the following form:

\begin{equation}
\mathcal{N}_u\left(\bm{\psi_u}; \bm{u}_{h}^{n}, p_{h}^{n} \right) = 0, \quad\forall \bm{\psi_u} \in \mathcal{U}_{h}^{\mathrm{CG}_{2}}\left(\mathcal{T}_{h}\right),
\label{eq:N_u}
\end{equation}
at each time step $t^n$, where 

\begin{equation*}
\begin{split}
\mathcal{N}_u\left(\bm{\psi_u}; \bm{u}_{h}^{n}, p_{h}^{n} \right) = &\sum_{T \in \mathcal{T}_{h}} \int_{T} \boldsymbol{\sigma}^{\prime}\left(\bm{u}_{h}^{n}\right) : \nabla^{s} \bm{\psi_u} \: d V  -  \sum_{T \in \mathcal{T}_{h}} \int_{T} \alpha  p_{h}^{n} \mathbf{I}  : \nabla^{s} \bm{\psi_u} \: d V\\
&-\sum_{T \in \mathcal{T}_{h}} \int_{T} \bm{f} \bm{\psi_u} \: d V-\sum_{e \in \mathcal{E}_{h}^{N,u}} \int_{e} \bm{t_{D}} \bm{\psi_u} \: d S, \quad \forall \bm{\psi_u} \in \mathcal{U}_{h}^{\mathrm{CG}_{2}}\left(\mathcal{T}_{h}\right)
\end{split}
\end{equation*}

\noindent 
Here $\int_{T} \cdot\  d V$ and $\int_{e} \cdot\  d S$ refer to volume and surface integrals, respectively, and $\nabla^{s}$ is the symmetric gradient operator. Furthermore, the notation for $\mathcal{N}_u\left(\bm{\psi_u}; \bm{u}_{h}^{n}, p_{h}^{n} \right)$ in \eqref{eq:N_u} highlights before the semicolon the test function, and after the semicolon the (possibly nonlinear) dependence on discrete solutions to the coupled problem. The same notation will be used hereafter for the remaining equations.

Next, the weak form of the mass balance equation \eqref{eq:mass_balance} is obtained multiplying by $\psi_p \in \mathcal{P}_{h}^{\mathrm{DG}_{0}}\left(\mathcal{T}_{h}\right)$ and integrating by parts, resulting in:

\begin{equation}
\mathcal{N}_p\left(\psi_p;  p_{h}^{n},\bm{q}_{h}^{n},c_{h}^{n}  \right) = 0, \quad\forall \psi_p \in \mathcal{P}_{h}^{\mathrm{DG}_{0}}\left(\mathcal{T}_{h}\right),
\label{eq:N_p_3f}
\end{equation}

\noindent
for each time step $t^n$, where 

\begin{equation*}
\begin{split}
\mathcal{N}_p\left(\psi_p; p_{h}^{n},\bm{q}_{h}^{n},c_{h}^{n} \right) & = 
\sum_{T \in \mathcal{T}_{h}} \int_{T}  \left(\frac{1}{M}+\frac{\alpha^{2}}{K}\right) 
\mathrm{BDF}_{1}\left( p_{h}^{n} \right) \psi_p \: d V
+ \sum_{T \in \mathcal{T}_{h}} \int_{T} \nabla \cdot \left( \bm{q}_{h}^{n} \right) \psi_{p} \: d V\\
& + \sum_{T \in \mathcal{T}_{h}} \int_{T} \frac{\alpha}{K} \mathrm{RK}_1(\sigma_{v}) \psi_p \: d V  + \sum_{T \in \mathcal{T}_{h}} \int_{T} \mathrm{RK}_1( \phi_c) \psi_p \: d V
- \sum_{T \in \mathcal{T}_{h}} \int_{T} g \psi_{p} \: d V.
\end{split}
\end{equation*}

\noindent
For the Darcy velocity equation \eqref{eq:darcy}, we obtain
\begin{equation}
\mathcal{N}_v\left(\bm{\psi}_{v}; \bm{u}_{h}^{n}, p_{h}^{n}, \bm{q}_{h}^{n},  c_{h}^{n} \right) = 0, \quad\forall \bm{\psi}_{v} \in \mathcal{V}_{h}^{\mathrm{BDM}_{1}}\left(\mathcal{T}_{h}\right).
\label{eq:N_v_3f}
\end{equation}

\noindent
where

\begin{equation*}
\begin{split}
\mathcal{N}_v\left(\bm{\psi}_{v}; \bm{u}_{h}^{n}, p_{h}^{n}, \bm{q}_{h}^{n},   c_{h}^{n} \right) & := \sum_{T \in \mathcal{T}_{h}} \int_{T} p_{h}^{n} \nabla \cdot \bm{\psi_v} \: d V \\ &+\sum_{T \in \mathcal{T}_{h}} \int_{T} \bm{k}(\bm{u}_{h}^{n}, c_{h}^{n})^{-1} \mu(c_{h}^{n}) \bm{q}_{h}^{n} \bm{\psi_v} \: d V\\
&+ \sum_{e \in \mathcal{E}_{h}^{D,m}} \int_{e} p_{D} \bm{\psi}_{v} \cdot \mathbf{n} \: d S.
\end{split}
\end{equation*}

\noindent

Lastly, for the advection-diffusion-reaction equations of species transport we write:

\begin{equation}
\mathcal{N}_c\left(\psi_c ; \bm{u}_{h}^{n} ,\bm{q}_{h}^{n},c_{h}^{n} \right) = 0, \quad\forall \psi_c \in \mathcal{P}_{h}^{\mathrm{EG}_{1}}\left(\mathcal{T}_{h}\right)
\label{eq:N_c}
\end{equation}

\noindent
for each time step $t^n$, where 
\begin{equation*}
\begin{split}
\mathcal{N}_c\left(\psi_c ; \bm{u}_{h}^{n} ,\bm{q}_{h}^{n},c_{h}^{n} \right) & = \sum_{T \in \mathcal{T}_{h}} \int_{T} \phi \mathrm{BDF}_{4}\left( c_{h}^{n} \right) \psi_c \: d V
+ \sum_{T \in \mathcal{T}_{h}} \int_{T} {\bm{D}_{e}}^*(\phi^n)\nabla c_h^n \cdot \nabla \psi_{c} \: d V \\ & - \sum_{e \in \mathcal{E}_h^{I} }  \int_{e}\left\{{\bm{D}_{e}}^*(\phi^n)\nabla c_h^n\right\}_{\delta_{e}} \cdot \llbracket \psi_c \rrbracket \: d S \\ & + \theta\sum_{e \in \mathcal{E}_h^{I} }  \int_{e}\left\{{\bm{D}_{e}}^*(\phi^n)  \nabla \psi_{c}\right\}_{\delta_{e}} \cdot \llbracket c_h^n \rrbracket \: d S \\ & +  \sum_{e \in \mathcal{E}_h^{I} }  \int_{e} \frac{\beta}{h_{e}} {{\bm{D}_{e}}^*(\phi^n)}_{{e}}  \llbracket c_h^n \rrbracket \cdot \llbracket \psi_c \rrbracket \: d S \\
&-\sum_{T \in \mathcal{T}_{h}} \int_{T} \bm{q}_h^n  c_{h}^{n} \cdot \nabla \psi_c \: d V +\sum_{e \in \mathcal{E}_h^{I} }  \int_{e} \bm{q}_h^n \cdot \mathbf{n} c^{\mathrm{up}}_h\llbracket \psi_c\rrbracket \: d S \\
&+\sum_{e \in \mathcal{E}_{h}^{\mathrm{Out}}} \int_{e} \bm{q}_h^n  \cdot \mathbf{n} c_{h}^{n} \psi_c \: d S\\
&-\sum_{T \in \mathcal{T}_{h}} \int_{T} R_{c} A_{s} \psi_{c} \: d V+\sum_{e \in \mathcal{E}_{h}^{\mathrm{In}}} \int_{e} \bm{q}_h^n \cdot \mathbf{n} c_{\mathrm{in}} \psi_c \: d S.
\end{split}
\end{equation*}

We note that the ${\bm{D}_{e}}^*(\phi^n)$ is redefining ${\bm{D}_{e}}(\phi^n)$   by including the numerical stabilization term, where 
\begin{equation}\label{eq:stab}
{\bm{D}_{e}}^*(\phi^n) := \bm{D}_{e}(\phi^n) + \gamma h \left\|\bm{q}_h^{n}\right\| \mathbf{I},
\end{equation}
as defined in \cite{araya2005adaptive,harari1994stabilized,masud2004multiscale}. 
The $ \gamma h \left\|\bm{q}_h^{n}\right\| \mathbf{I}$ term is often referred as the first order artificial diffusivity coefficient \cite{onate1998derivation,brezzi1992relationship}.  In our paper, we set the tuning parameter $\gamma=0.25$.
Alternative stabilization strategies including streamline diffusion and crosswind diffusion, or entropy viscosity methods could be also utilized to reduce oscillations in the numerical solution to the concentration field \cite{bonito2014stability,araya2005adaptive,harari1994stabilized,masud2004multiscale,brezzi1992relationship,GUERMOND2019143,scovazzi2017analytical,lee2017adaptive}.

Also, $c^{\mathrm{up}}_h$ is an upwind value of $c_{h}^{n}$ defined as \cite{riviere2000discontinuous,riviere2008discontinuous}:
\begin{equation}
c^{\mathrm{up}}_h=\left\{\begin{array}{ll}
c_{h}^{n+} & \text { if } \quad \bm{q}_h^n \cdot \mathbf{n} \geq 0 \\
c_{h}^{n-} & \text { if } \quad \bm{q}_h^n \cdot \mathbf{n}<0
\end{array} \quad \forall e=\partial T^+ \cap \partial T^-\right.
\end{equation}
\noindent
where $c_{h}^{n+}$ and $c_{h}^{n-}$ correspond to $c_{h}^{n}$ of $T^+$ and $T^-$, respectively.

Lastly, the two parameters $\theta$ and $\beta$ define corresponding interior penalty methods. 
The discretization becomes the symmetric interior penalty Galerkin method (SIPG) when $\theta = -1$, 
the incomplete interior penalty Galerkin method (IIPG)  when $\theta = 0$, 
and the non-symmetric interior penalty Galerkin method (NIPG) when $\theta = 1$ \cite{riviere2008discontinuous}. In this study, we set $\theta = -1$ for the simplicity and $\beta =1.1$ throughout this paper.

\begin{remark}
For the momentum balance equation \eqref{eq:linear_balance}, the traction boundary condition $\bm{t_D}$ (traction) is applied weakly on each $e \in \mathcal{E}_{h}^{N,u}$ in \eqref{eq:N_u}, while the displacement boundary condition $\bm{u}_D$ is strongly enforced on each $e \in \mathcal{E}_{h}^{D,u}$.
For the mass balance equation \eqref{eq:mass_balance}, since we use a mixed formulation, the flux boundary condition ${q_D}$ is strongly applied on each $e \in \mathcal{E}_{h}^{N,m}$, but the pressure boundary condition $p_D$ is weakly applied on each $\mathcal{E}_{h}^{D,m}$ in \eqref{eq:N_v_3f}. Finally, for the transport equation \eqref{eq:transport}, all boundary conditions are weakly applied in \eqref{eq:N_c}.
\end{remark}

\begin{remark}

In our computational framework, we provide a flexible choice of the time discretization schemes for each equation.
We use $\mathrm{BDF}_{1}$ for the time discretization of the mass balance equation \eqref{eq:mass_balance} since it is sufficient to provide the optimal error convergence rate, see \cite{zhang2016mixed}. 
For the time discretization of the transport equation \eqref{eq:transport}, 
we use $\mathrm{BDF}_{4}$ to capture a sharp front in the advection dominated regime \cite{riviere2008discontinuous}. 
\end{remark}

\subsection{Splitting algorithm}\label{sec:split}
The coupled system obtained from the discrete governing equations \eqref{eq:N_u}, \eqref{eq:N_p_3f}, \eqref{eq:N_v_3f}, and \eqref{eq:N_c} is nonlinear. 
Although the coupled nonlinear system may be solved in a monolithic manner, here we focus on developing a splitting algorithm for sequential solution to the coupled system, which can provide more flexibility especially when different software packages need to be combined.
The overall computational strategy is summarized in Algorithm \ref{al:1}.

\algnewcommand\algorithmicforeach{\textbf{for each}}
\algdef{S}[FOR]{ForEach}[1]{\algorithmicforeach\ #1\ \algorithmicdo}

\begin{algorithm}[H]
\caption{Splitting algorithm for hydro-mechanical-chemical coupling model}\label{al:1}
\begin{algorithmic}[1]
\State Initialize all input parameters \Comment{$p_0$ and $c_0$ must be provided.}
\State Solve the equilibrium state for $\bm{u_0}$ \Comment{see \eqref{eq:N_u}}
\State Update $\phi^{0}$, $\bm{k}^{0}$, $\bm{D}_{e}^{0}$, and $A_s^{0}$ \Comment{see \eqref{eq:porosity_by_mech}, \eqref{eq:perm_update_rut}, \eqref{eq:D_update}, \eqref{eq:a_s_update}}
\ForEach {time step $t^n$}
\State \emph{Part 1: coupling solid and fluid mechanics}
\State Set $\iota \rightarrow 0$  as the nonlinear iterations counter
\State ${p}_h^{n-1} \rightarrow p_h^{n,\iota=0}$, $\bm{q}_h^{n-1} \rightarrow \bm{q}_h^{n,\iota=0}$, $\bm{u}_h^{n-1} \rightarrow \bm{u}_h^{n,\iota=0}$ 
\ForEach {fixed stress iteration step $\left( \cdot \right)^{n, \iota}$ until $\delta \phi^{n,\iota}$ $<$ $\mathrm{TOL}$} \label{line:fixed_stress}
\State Solve \eqref{eq:N_p_3f} and \eqref{eq:N_v_3f} w.r.t. $p_h^{n}$ and $\bm{q}_h^{n}$ for fixed $\bm{u}_h^{n} := \bm{u}_h^{n,\iota-1}, c_h^{n} :=\hat{c}_h^{n}$ to get $p_h^{n, \iota}, \bm{q}_h^{n,\iota}$  \label{line:pressure_velocity}
\State Calculate $\phi_f^{n,\iota}$  \Comment{see \eqref{eq:porosity_by_flow}}
\State Solve \eqref{eq:N_u} w.r.t. $\bm{u}_h^{n}$ and for fixed $p_h^{n} := p_h^{n, \iota}$ to get $\bm{u}_h^{n,\iota}$ \label{line:displacement}
\State Calculate $\phi_m^{n,\iota}$  \Comment{see \eqref{eq:porosity_by_mech}}
\State Evaluate $\mathrm{F}\left(\Dot{\sigma_{v}}^{n,\iota} \right)$ \Comment{see \eqref{eq:F_sigma}} \label{line:evalute_stress}
\State Evaluate $\delta \phi^{n,\iota}$ \Comment{see \eqref{eq:del_phi}}
\State Update $\bm{k}^{n,\iota}$ \Comment{see \eqref{eq:porosity_by_mech}, \eqref{eq:perm_update_rut}}
\EndFor 
\State ${p}_h^{n,\iota} \rightarrow {p}_h^{n}$, $\bm{q}_h^{n,\iota} \rightarrow \bm{q}_h^{n}$, $\bm{u}_h^{n,\iota} \rightarrow \bm{u}_h^{n}$ \label{line:fixed_stress_end}
\State \emph{Part 2: chemical process}\label{line:after_fixed_stress}
\State Update $\phi^{n}$, $\bm{D}_{e}^{n}$, and $A_s^{n}$ \Comment{see \eqref{eq:porosity_by_mech}, \eqref{eq:D_update}, \eqref{eq:a_s_update}}\label{line:update_para_after_fixed_stress}
\State Solve \eqref{eq:N_c} w.r.t. ${c}_h^{n}$ for fixed $\bm{u}_h^{n} := \bm{u}_h^{n}, \bm{q}_h^{n} := \bm{q}_h^{n}$ to get ${c}_h^{n}$
\State Extrapolate $\hat{c}_h^{n+1}$ \Comment{see \eqref{eq:c_extrapolate}} \label{line:extra}
\State Calculate $\Delta t^{n+1}$ \Comment{see \eqref{eq:time_mult}}
\State Evaluate $\mathrm{F}\left(\hat{q}^{n+1} \right)$ and $\mathrm{F}\left(\hat{\Dot{\phi_{c}}}^{n+1} \right)$ \Comment{see \eqref{eq:F_q}, \eqref{eq:F_phi}}
\State Update $\hat{\phi}^{n+1}$, $\hat{\bm{k}}^{n+1}$, $\hat{\bm{D}}_t^{n+1}$, $\hat{\mu}^{n+1}$, and $\hat{A_s}^{n+1}$ \Comment{see \eqref{eq:c_extrapolate}, \eqref{eq:porosity_by_chem}, \eqref{eq:perm_update_rut}, \eqref{eq:D_update}, \eqref{eq:mu_update}, \eqref{eq:a_s_update}} \label{line:porosity_by_chem}
\State $p_h^{n} \rightarrow p_h^{n-1}$, $\bm{q}_h^{n} \rightarrow \bm{q}_h^{n-1}$, $\bm{u}_h^{n} \rightarrow \bm{u}_h^{n-1}$, $c_h^{n} \rightarrow c_h^{n-1}$ \Comment{update time step $n-1$}
\State Output
\EndFor
\end{algorithmic}
\end{algorithm}

In Algorithm \ref{al:1}, we separate our algorithm into two parts. The first part (lines \ref{line:fixed_stress} to \ref{line:fixed_stress_end}) 
focuses on solving the coupled hydro-mechanical problem, \eqref{eq:N_u}, \eqref{eq:N_p_3f}, and \eqref{eq:N_v_3f}, using the fixed stress method which is an unconditionally stable splitting scheme\cite{kim2011stability, dana2018convergence,dana2018multiscale,mikelic2013convergence}. 
At each iteration $\iota$ we solve \eqref{eq:N_p_3f} and \eqref{eq:N_v_3f} for the velocity
$\bm{q}_h^{n,\iota}$ and the pressure ${p}_h^{n,\iota}$ using a monolithic method (line \ref{line:pressure_velocity}) based on given displacement $\bm{u}_h^{n,\iota-1}$ from previous nonlinear iteration and concentration extrapolated $\hat{c}_h^{n}$ from previous time step. 
Then, we couple with \eqref{eq:N_u} using the fixed-stress split scheme based on the pressure  ${p}_h^{n,\iota}$ computed at the current nonlinear iteration (line \ref{line:displacement}). 
The convergence criterion is based on $\delta \phi^{n,\iota}$ (Algorithm \ref{al:1} line \ref{line:fixed_stress}), which is defined as:


\begin{equation}\label{eq:del_phi}
\delta \phi^{n,\iota} := \frac{\phi_m^{n,\iota} - \phi_f^{n,\iota}}{\phi_m^{n,\iota}}.
\end{equation}

\noindent
Here, $\phi_m^{n,\iota}$ is the porosity resulting from the solid deformation \eqref{eq:porosity_by_mech} and $\phi_f^{n,\iota}$ is the porosity resulting from the fluid flow problem defined as \cite{mikelic2013convergence,dana2018convergence,dana2018multiscale}:

\begin{equation}\label{eq:porosity_by_flow}
\phi_f^{n,\iota}=\phi^{n-1}+\frac{\left(\alpha-\phi^{n-1}\right)}{K} \left( p^{n,\iota} - p^{n-1} \right),
\end{equation}

\noindent
where $\left( \cdot \right)^\iota$ represents iteration counter inside the fixed-stress loop. From the fixed stress split concept \eqref{eq:porosity_by_flow} is the $\phi$ predictor, while \eqref{eq:porosity_by_mech} is the $\phi$ corrector \cite{kim2011stability, dana2018convergence,dana2018multiscale,mikelic2013convergence}. Hence, when $\phi_m^{n,\iota}$ and $\phi_f^{n,\iota}$ converge, i.e., $\delta \phi^{n,\iota}$ $<$ $\mathrm{TOL}$, the fixed-stress loop is completed. 
The tolerance $\mathrm{TOL}$ is set as $1\times10^{-6}$ throughout this study. Note that the flow equations, \eqref{eq:N_p_3f} and \eqref{eq:N_v_3f}, are solved by assuming that $\frac{\partial \sigma_{v}}{\partial t} = 0$, i.e., $\mathrm{F}\left(\Dot{\sigma_{v}}^{n,\iota}  \right)$ is frozen; therefore, this term is evaluated explicitly after the momentum equation \eqref{eq:N_u} is solved, as illustrated in Algorithm \ref{al:1} line \ref{line:evalute_stress} \cite{kim2011stability, mikelic2013convergence}, and $\mathrm{F}\left(\Dot{\sigma_{v}}^{n,\iota} \right)$ is defined as:

\begin{equation}\label{eq:F_sigma}
\mathrm{F}\left(\Dot{\sigma_{v}}^{n,\iota}  \right) := \sum_{T \in \mathcal{T}_{h}} \int_{T} \frac{\alpha}{K} \mathrm{RK}_1 \left(\sigma_{v} (\bm{u}^{n,\iota}, \bm{u}^{n-1},{p}^{n,\iota}, {p}^{n-1})\right) \psi_p \: d V.
\end{equation}

The second part (from line \ref{line:after_fixed_stress}) focuses on solving advection-diffusion-reaction equation \eqref{eq:N_c}, using $\bm{q}_h^{n}$, $\phi^{n}$, $\bm{D}_{e}^{n}$, and $A_s^{n}$ obtained from the first part.
One could view this strategy as a one-way coupling scheme between coupled hydro-mechanical and advection-diffusion-reaction equations.
Next, Algorithm \ref{al:1} line \ref{line:extra} linearizes ${c}_{h}^{n+1}$ by extrapolating ${c}_{h}^{n}$ and ${c}_{h}^{n-1}$ to $\hat{c}_{h}^{n+1}$ by using \eqref{eq:gen_extrapolate}:

\begin{equation} \label{eq:c_extrapolate}
\hat{c}_{h}^{n+1} = \mathrm{EX}\left( {c}_{h} \right)
\end{equation}

\noindent
where $\left( \hat{\cdot} \right)^n$ represents an extrapolation value based on the extrapolation described in \eqref{eq:gen_extrapolate}.
Subsequently, we evaluate $\mathrm{F}\left(\Dot{\phi_{c}}^{n,\iota} \right)$ and $\mathrm{F}\left(q^{n,\iota} \right)$, which are defined as 

\begin{equation}\label{eq:F_phi}
\mathrm{F}\left(\hat{\Dot{\phi_{c}}}^{n+1} \right) := \sum_{T \in \mathcal{T}_{h}} \int_{T} \mathrm{RK}_1 \left( \phi_c (\hat{c}^{n+1}, {c}^{n}) \right) \psi_p \: d V,
\end{equation}

\noindent
and

\begin{equation}\label{eq:F_q}
\mathrm{F}\left(\hat{q}^{n+1} \right) := \sum_{T \in \mathcal{T}_{h}} \int_{T} R_{c}(\hat{c}^{n+1},p^{n}) A_{s} (\hat{c}^{n+1},\bm{u}^{n})  \psi_{c} \: d V,
\end{equation}

\noindent
using $\hat{c}_{h}^{n+1}$ calculated by \eqref{eq:c_extrapolate}. 
We note that the equation \eqref{eq:N_c} becomes linear by employing $\hat{c}_{h}^{n+1}$ to calculate $\mathrm{F}\left(q \right)$. Also, the porosity alteration as a result of calcite dissolution/precipitation (Algorithm \ref{al:1} line \ref{line:porosity_by_chem}) is computed by

\begin{equation}\begin{aligned}
\hat{\phi}^{n+1} = \hat{\phi}_c^{n+1} &= \mathrm{RK}_4 \left( \Gamma \left(\bm{u}^{n}, \hat{c}_{h}^{n+1}\right) \right). \\
\end{aligned}\end{equation}

\noindent
Note that the porosity change due to the calcite dissolution/precipitation reactions is  additional to the porosity change by solid deformation, \eqref{eq:porosity_by_mech}. Subsequently, $\hat{\bm{k}}^{n+1}$, $\hat{\bm{D}}_t^{n+1}$, and $\hat{A_s}^{n+1}$ are determined using $\hat{\phi}^{n+1}$. Lastly, we also calculate $\hat{\mu}^{n+1}$ using $\hat{c}_{h}^{n+1}$, see \eqref{eq:c_extrapolate} and \eqref{eq:mu_update}.

For all the computations, matrices and vectors are built using the FEniCS form compiler \cite{AlnaesBlechta2015a}. The block structure is assembled by using the multiphenics toolbox \cite{Ballarin2019}. Solvers are employed from PETSc package \cite{petsc-user-ref}. All simulations are computed on $\mathrm{XeonE5\_2650v4}$ with a single thread.

\begin{remark}
 We note that the EG method, which is used to approximate the advection-diffusion-reaction \eqref{eq:transport}, is based on the Galerkin method, which could be extended to consider adaptive meshes that contain hanging nodes. Besides, an adaptive enrichment, i.e., the piecewise-constant functions only added to the elements where the sharp material discontinuities are observed, can be developed. 
\end{remark}

\section{Numerical examples}\label{sec:results}

\begin{figure}[!ht]
   \centering
        \includegraphics[keepaspectratio, height=9.0cm]{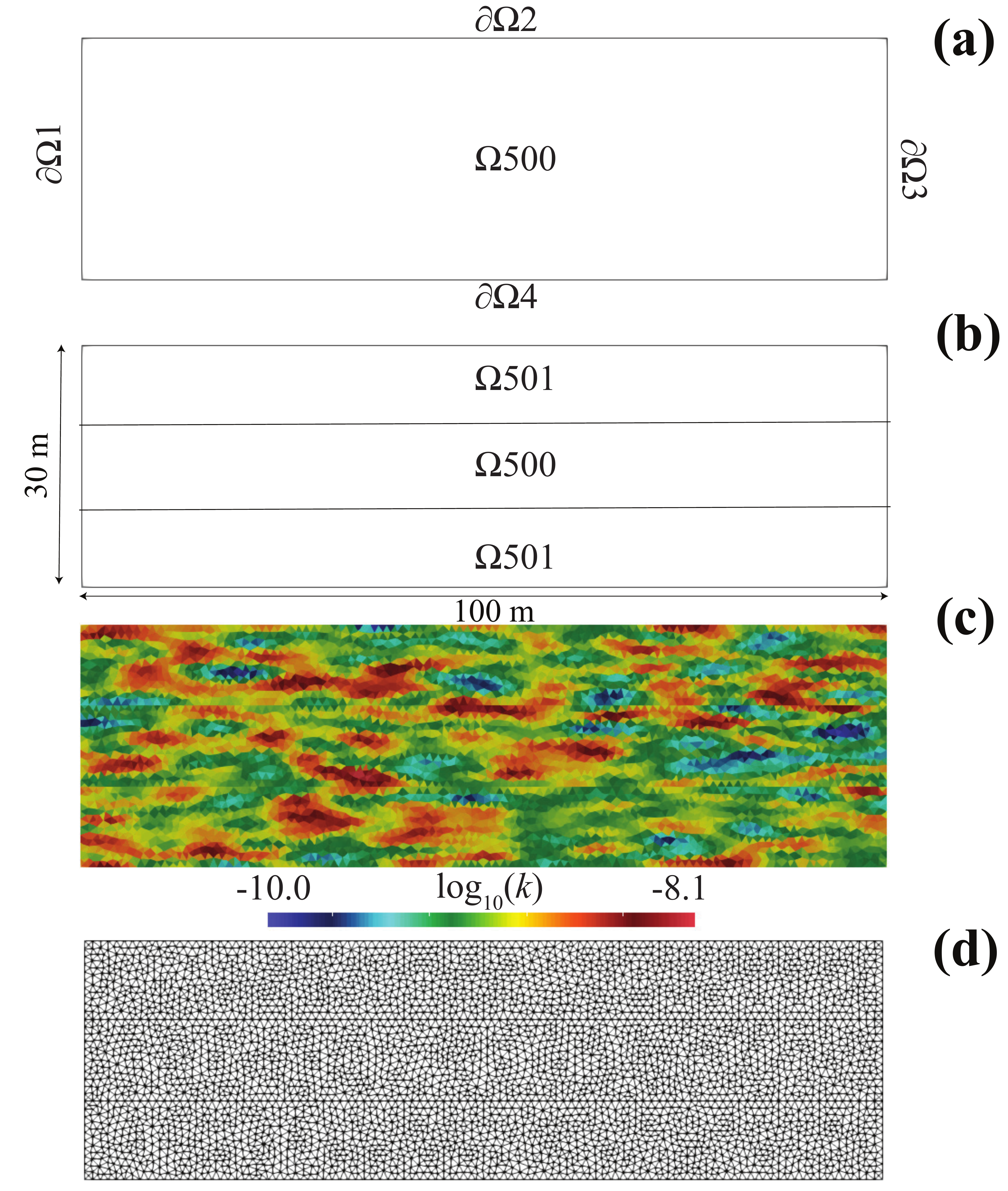}
   \caption{Geometry and notation used to define material properties; ($\bm{\mathrm{a}}$) example 1: single-layer porous medium ($\Omega_{500}$), ($\bm{\mathrm{b}}$) example 2: three-layer porous medium ($\Omega_{500}$ and $\Omega_{501}$), ($\bm{\mathrm{c}}$) example 3: heterogeneous porous medium (the arithmetic mean of $\bm{k} = 8.8 \times 10^{-10} \mathbf{I}$ \si{m^2} with correlation length in $x$- and $y$-direction of 5 and 1 m, respectively.), and ($\bm{\mathrm{d}}$) mesh used for all examples (number of element is 7852).}
   \label{fig:hmc_2d_cag}
\end{figure}

In this section, we demonstrate the performance and capabilities of the proposed numerical method through various numerical examples. 
We begin with a single-layer model comparing the performance for single-phase flow with chemical dissolution/precipitation and solid deformation. Then we illustrate the performance of the developed model for a layered medium as well as a heterogeneous single-layer medium. Lastly, we test the proposed framework using an example with an anisotropic permeability field.  All four examples and their mesh are illustrated in Figure \ref{fig:hmc_2d_cag}. More detailed setup, including the input parameters and the boundary conditions of each example, are described in the beginning of each example.

\subsection{Example 1}

In the first example, the computational domain is defined as $\Omega_{500}$ = $[0, 100] \times [0, 30]$, which presents a single layer as shown in Figure \ref{fig:hmc_2d_cag}a.
Following the typical physical properties of rocks \cite{jaeger2009fundamentals}, we set $K=8.4$ \si{GPa}, $\alpha = 0.74$, $\nu = 0.18$, $\phi = 0.2$, $\bm{k} = 8.8 \times 10^{-10} \mathbf{I}$ \si{m^2}. The fluid properties considered in this case are $c_f = 1.0 \times 10^{-10}$ \si{Pa^{-1}}, $\rho = 1000$ \si{kg/m^3}, $\bm{D} = 1.0 \times 10^{-12}$ \si{m^2/s}, and $\mu$ is calculated using \eqref{eq:mu_update} by setting $\mu_h = 5.0$ and $\mu_l = 1.0 \times 10^{-4}$ \si{Pa/s} corresponding to $c_h = 1.68$ and $c_l = 0.0$ \si{mmol/m^3}, respectively. 
Next, the boundary conditions for all these examples are applied as follows. 
For the momentum balance equation \eqref{eq:linear_balance}, we assume $\bm{u}_D \cdot \mathbf{n} = 0$ on $\partial \Omega_{1}$, $\partial \Omega_{3}$, and $\partial \Omega_{4}$. Furthermore, $\bm{t}_D = \left[ 0.0, -2.0 \times 10^{6} \right]$ \si{Pa} is applied on $\partial \Omega_{2}$. Therefore, the medium is under compression. 
For the mass balance equation \eqref{eq:mass_balance}, the boundary condition $q_D = 0$ is set on $\partial \Omega_{2}$ and $\partial \Omega_{4}$ and we impose $p_D = 1 \times 10^{5}$ \si{Pa} on $\partial \Omega_{3}$.
Here, for the mass balance equation \eqref{eq:mass_balance}, we test two different scenarios on $\partial \Omega_{1}$, where scenario (a) corresponds to $q_D = 2 \times 10^{-4}$ \si{m/s} and scenario (b) is characterized by $q_D = 1 \times 10^{-4}$ \si{m/s}.
Thus, scenarios (a) and (b) will be referred to as the high and low injection rate cases, respectively.
Since we want to compare the results of the above scenarios at the same total injected volume (I.V.), which is defined as 

\begin{equation}
    \mathrm{I.V.} = q_D t^n A_{\mathrm{d}},
\end{equation}

\noindent
where $A_{\mathrm{d}}=30 \si{m^2}$ is the surface area of $\partial \Omega_{1}$, the time $t^n$ of the scenario (b) is twice to scenario (a). For the advection-diffusion-reaction equation \eqref{eq:transport}, we impose the inflow condition $c_{in} = 0.5$ on $\partial \Omega_{1}$. 
The initial pressure $p_0$ is $1 \times 10^{6}$ \si{Pa}, the initial concentration $c_0$ is calculated by \eqref{eq:c_sat} using $p = p_0$ and $\tau = 20$ \si{C}, and the initial displacement $\bm{u}_0$ is calculated as stated in Algorithm \ref{al:1}. The penalty parameter ($\beta$) is set to be 1.1 for the EG method. 
The $\mathrm{CFL}$ is used as 0.1 for calculating $\Delta t^n$, see \eqref{eq:time_mult}.

Here, we compare the transient distribution of the concentration achieved with the developed HMC coupled numerical scheme in a homogeneous porous medium for two different injection rates. The aim is to illustrate the impact of different processes on the advance of the flow path and reactive solute transport. Initially, the composition of the pore fluid within the porous medium is in equilibrium with calcite. Note that $c_{eq}$ calculated by \eqref{eq:c_sat} is a function of temperature and pressure. In this example, assuming constant temperature, pressure deviates from the initial fluid pressure in time and space. The changes in the pressure field as a result of fluid injection on the left boundary and the fluid production on the right boundary varies the $c_{eq}$ resulting in precipitation or dissolution in the domain. The injected water is also unsaturated with respect to calcite. Therefore, the injected fluid, as advances into the domain, will dissolve the calcite mineral. 

\begin{figure}[!ht]
   \centering
        \includegraphics[width=16.0cm, keepaspectratio]{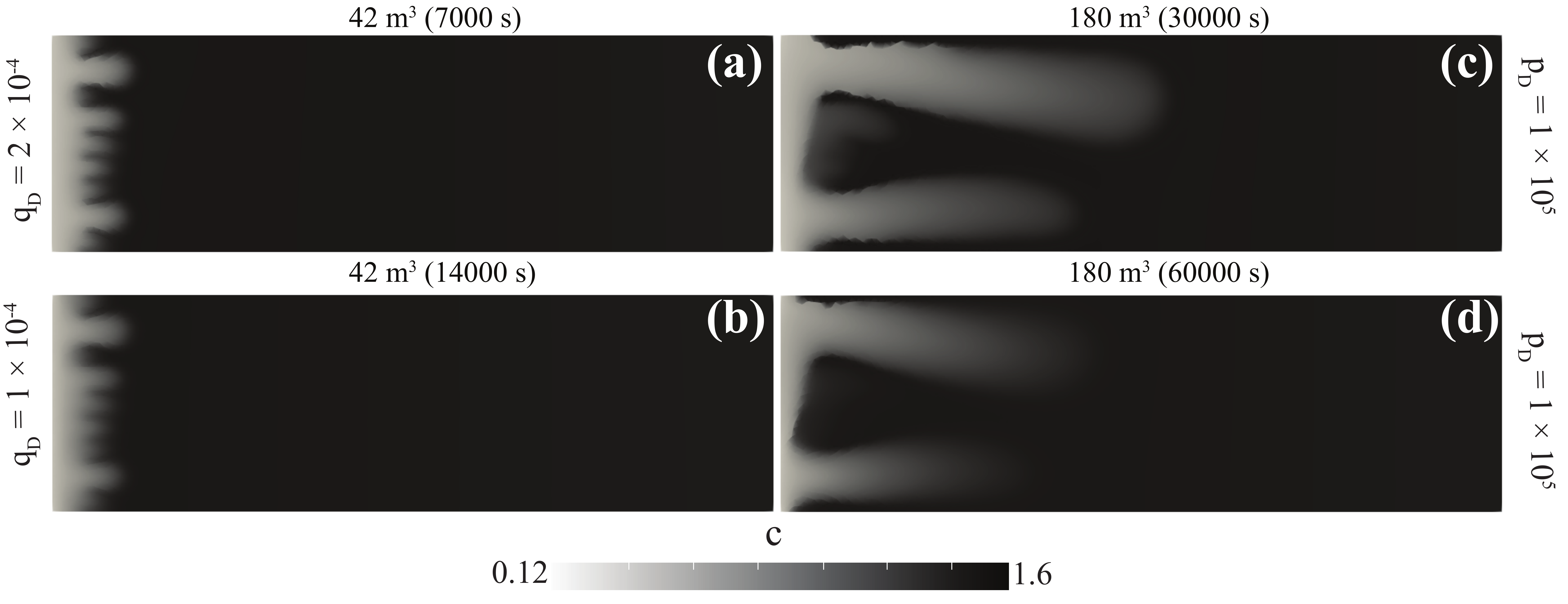}
   \caption{Example 1: concentration fields, $c$; $\mathrm{I.V.} = 42$ \si{m^3} using ($\bm{\mathrm{a}}$) $q_D = 2 \times 10^{-4}$ \si{m/s} at $\partial \Omega_{1}$ and ($\bm{\mathrm{b}}$) $q_D = 1 \times 10^{-4}$ \si{m/s} at $\partial \Omega_{1}$; and $\mathrm{I.V.} = 180$ \si{m^3} using ($\bm{\mathrm{c}}$) $q_D = 2 \times 10^{-4}$ \si{m/s} at $\partial \Omega_{1}$ and ($\bm{\mathrm{d}}$) $q_D = 1 \times 10^{-4}$ \si{m/s} at $\partial \Omega_{1}$. The boundary conditions shown in this picture are corresponding to $\partial \Omega_{1}$ and $\partial \Omega_{3}$ of the mass balance equation \eqref{eq:mass_balance}.}
   \label{fig:c_step1_hmc_2d_l1_cag}
\end{figure}

Figure \ref{fig:c_step1_hmc_2d_l1_cag} shows the concentration fields at different injected fluid volumes (I.V.) and for both scenarios associated to $q_D$. There are three main observations from these figures. The first one is the flow instability, or fingering, emerged as a result of the difference between the injected fluid viscosity and the in-situ fluid viscosity. The second observation is that for the higher injection rate scenario, the fingers are more developed at a later time compared to that of the low injection scenario.
The third one is that most of the fingers developed initially either merge or vanishes at the later stage, forming one or two main fingers. 

\begin{figure}[!ht]
   \centering
        \includegraphics[width=16.0cm, keepaspectratio]{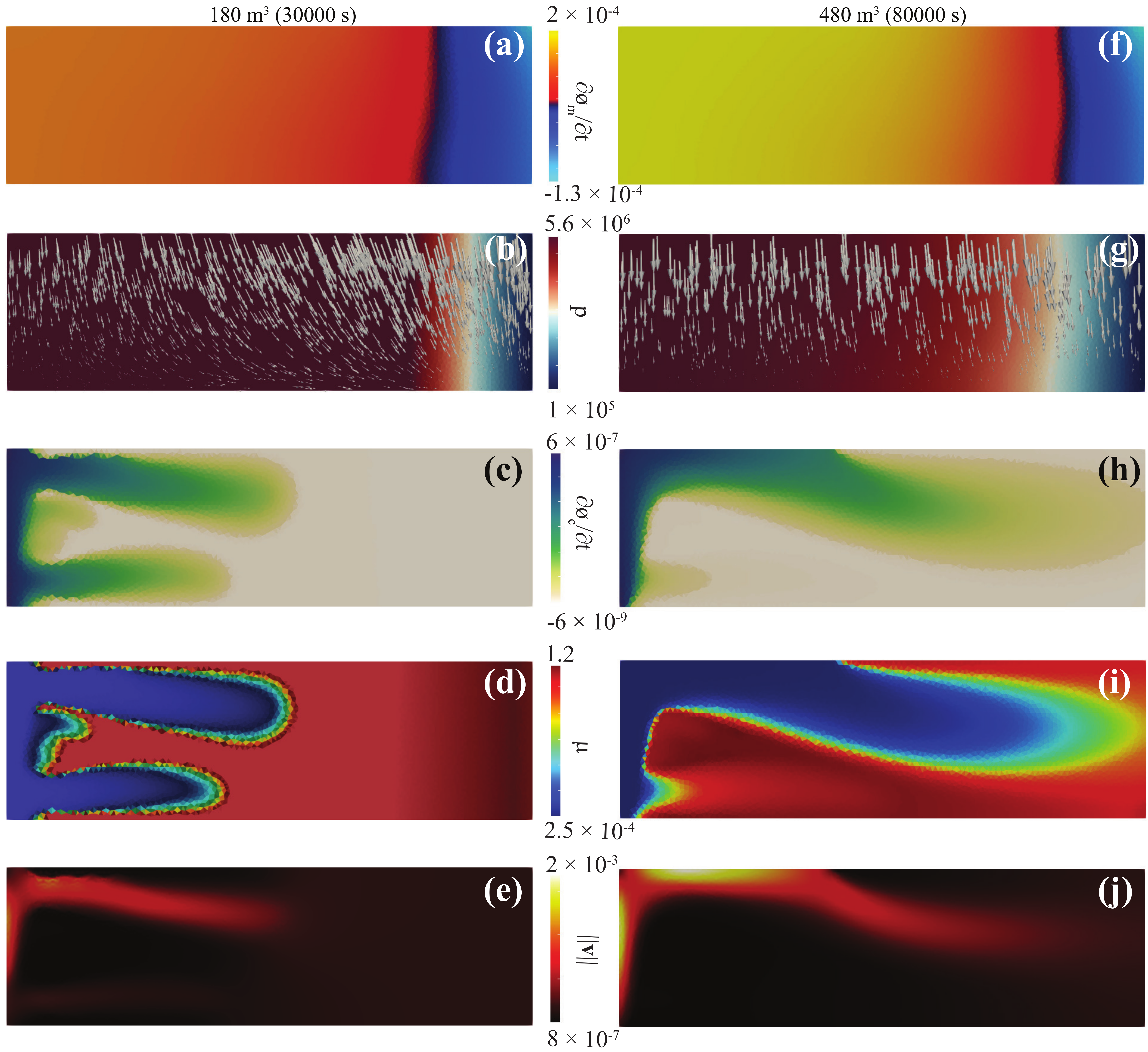}
   \caption{Example 1: using $q_D = 2 \times 10^{-4}$ \si{m/s} at $\partial \Omega_{1}$; $\mathrm{I.V.} = 42$ \si{m^3}: ($\bm{\mathrm{a}}$) the rate of change of porosity according to mechanics deformation, $\frac{\partial\phi_m}{\partial t}$, see \eqref{eq:porosity_by_mech}, ($\bm{\mathrm{b}}$) the fluid pressure, $p$, in surface and the displacement, $\bm{u}$, in grey arrows, ($\bm{\mathrm{c}}$) the rate of change of porosity according to calcite dissolution/precipitation, $\frac{\partial\phi_c}{\partial t}$, see \eqref{eq:porosity_by_chem}, ($\bm{\mathrm{d}}$) the fluid viscosity, $\mu$, and ($\bm{\mathrm{e}}$) the magnitude of the fluid velocity, $\left\|\bm{q}\right\|$. $\mathrm{I.V.} = 480$ \si{m^3}: ($\bm{\mathrm{f}}$) $\frac{\partial\phi_m}{\partial t}$, ($\bm{\mathrm{g}}$) $p$ in surface and $\bm{u}$ in grey arrows, ($\bm{\mathrm{h}}$) $\frac{\partial\phi_c}{\partial t}$, ($\bm{\mathrm{i}}$) $\mu$, and ($\bm{\mathrm{j}}$) $\left\|\bm{q}\right\|$.  Note that the magnitude of $\bm{u}$ is from $0.0$ to $3.0 \times 10^{-2}$, and the trend of the results of the $q_D = 1 \times 10^{-4}$ \si{m/s} at $\partial \Omega_{1}$ case are similar.}
   \label{fig:c_interplay_hmc_2d_l1_cag}
\end{figure}

Next, we present the interaction among different processes including mechanical deformation, calcite dissolution/precipitation, and viscosity alteration in Figure \ref{fig:c_interplay_hmc_2d_l1_cag} for two different time steps. Note that the results of the low injection rate case are similar (for the same volume of injected fluid); hence, we present here only the results of the high injection rate case. First, one could observe that the effect of mechanical deformation is dictated by both $p$ and $\bm{u}$, see Figure \ref{fig:c_interplay_hmc_2d_l1_cag}b and g. Figure \ref{fig:c_interplay_hmc_2d_l1_cag}a and f illustrate the reduction of $\phi$ by the solid deformation as the model is under compression. The increased fluid pressure by fluid injection, however, limits the porosity reduction. This is reflected in Figures \ref{fig:c_interplay_hmc_2d_l1_cag}a and f in which $\frac{\partial\phi_m}{\partial t}$ is positive in the left part of the domain and negative in the right part of the domain.

The $\frac{\partial\phi_c}{\partial t}$ result is shown in Figure \ref{fig:c_interplay_hmc_2d_l1_cag}c and h. Since the injected concentration $c_{in} = 0.5$ is lower than $c_0$ (initial $c_{eq}$), the porous medium is dissolved in places to which the injected fluid is transported. Note that $\frac{\partial\phi_c}{\partial t}$ is positive where the dissolution occurs and negative where the perception occurs. At this time step, the maximum magnitude of $\frac{\partial\phi_c}{\partial t}$ is $10^{-7}$, which is much less compared to that of $\frac{\partial\phi_m}{\partial t}$, which is around $10^{-4}$. We note this magnitude could be varied with different input parameters and boundary conditions of each equation, \eqref{eq:linear_balance}, \eqref{eq:mass_balance}, or \eqref{eq:transport}. The value of $\mu$ is also altered, see Figure \ref{fig:c_interplay_hmc_2d_l1_cag}d and i, as the concentration front progresses. This alteration causes the flow instability discussed previously and establishes a preferential flow path. The impact of $\frac{\partial\phi_m}{\partial t}$, $\frac{\partial\phi_c}{\partial t}$, and $\mu$ alteration can be seen in $\bm{q}$ field shown in Figure \ref{fig:c_interplay_hmc_2d_l1_cag}e and j. Interestingly, as the first finger reaches the outlet boundary the second finger gradually disappears resulting in only one preferential path between the inlet and outlet of the model.   

Thus, we have confirmed that the proposed framework can well simulate the expected physical and chemical phenomena including solid deformation, viscous fingering, and dissolution/precipitation. 
The key ingredients of this method are the capability for tracking the interface of the concentration species approximated by the high order methods with numerical stabilization, the computation of reaction terms with the EG method, and the locally conservative flux from BDM.

\subsection{Example 2}

In the second example, 
we consider three layers ($\Omega_{500}$=$[0, 100] \times [10, 20]$, $\Omega_{501}$=$[0, 100] \times [20, 30]$, and $\Omega_{502}$=$[0, 100] \times [0, 10]$) as the computational domain. See Figure \ref{fig:hmc_2d_cag}b.  
In $\Omega_{500}$, we set $\bm{k} = 8.8 \times 10^{-10} \mathbf{I}$ \si{m^2}, while $\bm{k} = 8.8 \times 10^{-11} \mathbf{I}$ \si{m^2} in $\Omega_{501}$ and {$\Omega_{502}$}. 
Thus, in this case, the top {$\Omega_{501}$} and bottom {$\Omega_{502}$} layers have one order of magnitude of $\bm{k}$ less than that of the middle layer {$\Omega_{500}$}.
All other rock and fluid parameters are the same as in the first example. 

\begin{figure}[!ht]
   \centering
        \includegraphics[width=16.0cm, keepaspectratio]{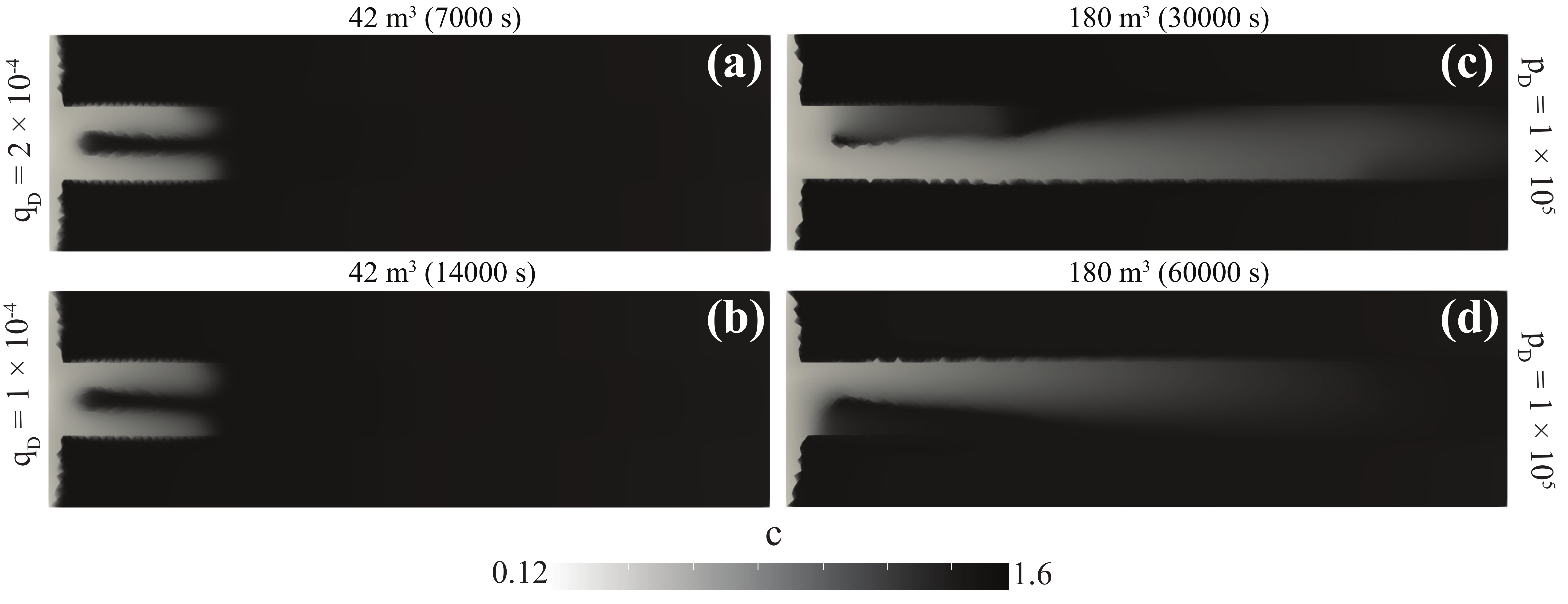}
   \caption{Example 2: concentration fields, $c$; $\mathrm{I.V.} = 42$ \si{m^3} using ($\bm{\mathrm{a}}$) $q_D = 2 \times 10^{-4}$ \si{m/s} at $\partial \Omega_{1}$ and ($\bm{\mathrm{b}}$) $q_D = 1 \times 10^{-4}$ \si{m/s} at $\partial \Omega_{1}$; and $\mathrm{I.V.} = 180$ \si{m^3} using ($\bm{\mathrm{c}}$) $q_D = 2 \times 10^{-4}$ \si{m/s} at $\partial \Omega_{1}$ and ($\bm{\mathrm{d}}$) $q_D = 1 \times 10^{-4}$ \si{m/s} at $\partial \Omega_{1}$. The boundary conditions shown in this picture are corresponding to $\partial \Omega_{1}$ and $\partial \Omega_{3}$ of the mass balance equation \eqref{eq:mass_balance}.}
   \label{fig:c_step1_hmc_2d_l3_cag}
\end{figure}

The concentration field $c$ for two different injection scenarios (as discussed in example 1) for the three-layer porous medium are presented in Figure \ref{fig:c_step1_hmc_2d_l3_cag}. Unlike the single-layer porous medium, even though the concentration fields at the early time are similar between the high, $q_D = 2 \times 10^{-4}$ \si{m/s}, and low, $q_D = 1 \times 10^{-4}$ \si{m/s}, injection rates, the progression of concentration field is different at the later time.
It appears that the dynamic of the coupled processes controlled by the injection rate can impact the development of the dominant finger in the middle layer. Note that since the top and bottom layers, $\Omega_{501}$ and $\Omega_{502}$, have lower permeability than the middle layer, $\Omega_{500}$, the flow mainly goes through the middle layer. 
Similar to the previous example, one of the two initial fingers becomes the main path connecting the inlet and the outlet boundaries. 

\begin{figure}[!ht]
   \centering
        \includegraphics[width=16.0cm, keepaspectratio]{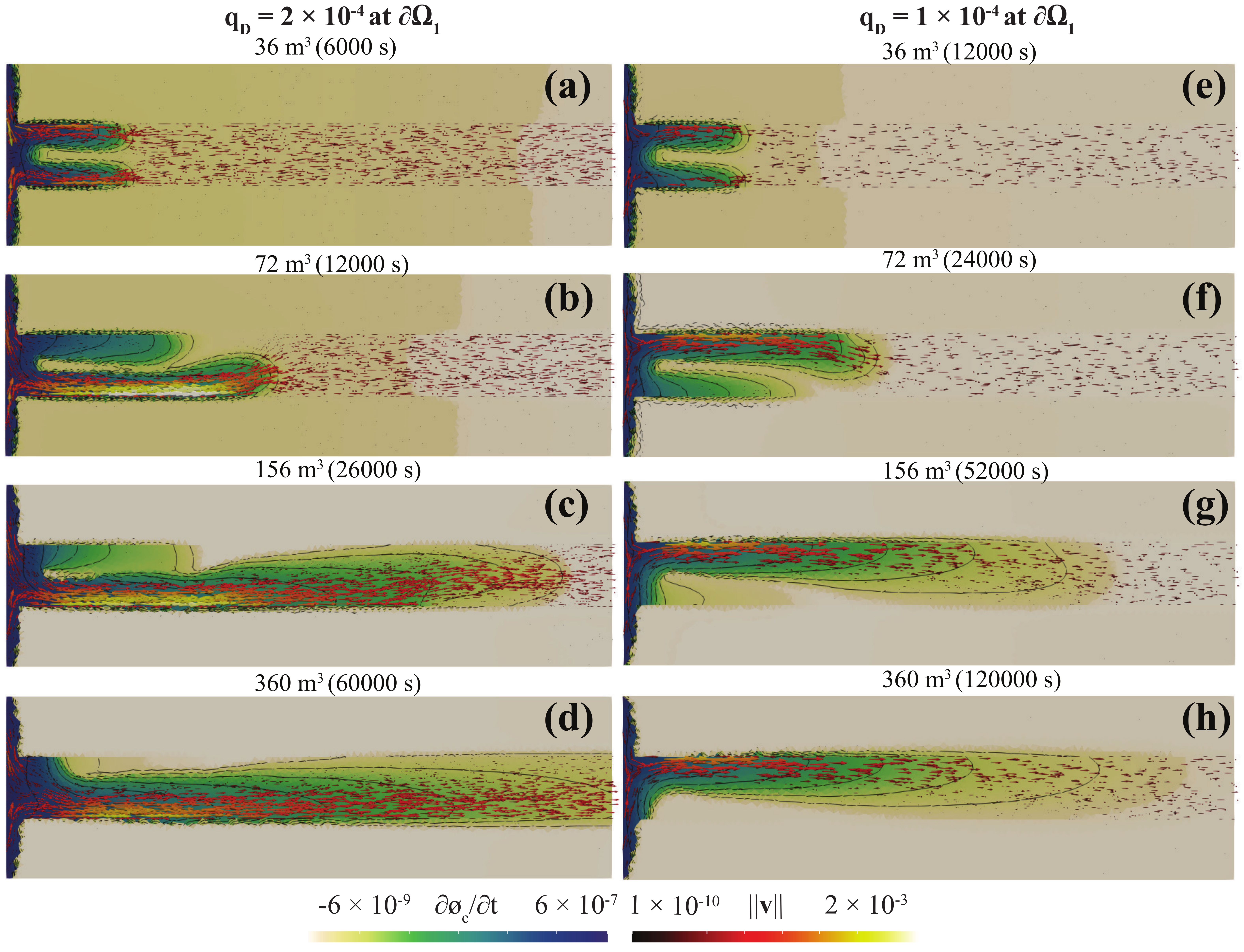}
   \caption{Example 2: the results of the rate of change of porosity according to calcite dissolution/precipitation ($\frac{\partial\phi_c}{\partial t}$) shown in surface plot, concentration ($c$), shown in black contour (10 contours ranging from 0.12 to 1.6), and the fluid velocity ($\bm{q}$) shown in arrows with $q_D = 2 \times 10^{-4}$ \si{m/s} at $\partial \Omega_{1}$; ($\bm{\mathrm{a}}$) $\mathrm{I.V.} = 36$ \si{m^3} ($t = 6000$ \si{s}), ($\bm{\mathrm{b}}$) $\mathrm{I.V.} = 72$ \si{m^3} ($t = 12000$ \si{s}), ($\bm{\mathrm{c}}$) $\mathrm{I.V.} = 156$ \si{m^3} ($t = 26000$ \si{s}), and ($\bm{\mathrm{d}}$) $\mathrm{I.V.} = 360$ \si{m^3} ($t = 60000$ \si{s}), and with $q_D = 1 \times 10^{-4}$ \si{m/s} at $\partial \Omega_{1}$; ($\bm{\mathrm{e}}$) $\mathrm{I.V.} = 36$ \si{m^3} ($t = 1200$ \si{s}), ($\bm{\mathrm{f}}$) $\mathrm{I.V.} = 72$ \si{m^3} ($t = 24000$ \si{s}), ($\bm{\mathrm{g}}$) $\mathrm{I.V.} = 156$ \si{m^3} ($t = 32000$ \si{s}), and ($\bm{\mathrm{h}}$) $\mathrm{I.V.} = 360$ \si{m^3} ($t = 120000$ \si{s}).}
   \label{fig:c_prog_high_rate_hmc_2d_l3_cag}
\end{figure}

In Figure \ref{fig:c_prog_high_rate_hmc_2d_l3_cag}, the behavior of the concentration and velocity fields, together with temporal porosity alteration ($\frac{\partial\phi_c}{\partial t}$), are illustrated for both injection scenarios.
As mentioned earlier, due to the difference of viscosity ($\mu$) between that of the injected $c$ and the in-situ $c$, two fingers developed at the beginning, see Figure \ref{fig:c_prog_high_rate_hmc_2d_l3_cag}a and e. For the high injection rate, the top finger, however, disappeared while the bottom finger progresses until it reaches the outlet $\partial \Omega_{3}$, see Figures \ref{fig:c_prog_high_rate_hmc_2d_l3_cag}b-d. One could see that the reaction front shown by $\frac{\partial\phi_c}{\partial t}$ progresses similarly to the concentration front shown by the black contours. Besides, as the concentration field develops, the change in $\mu$ enhances the flow channeling illustrated by velocity arrows. 
For the low injection rate case shown in Figure \ref{fig:c_prog_high_rate_hmc_2d_l3_cag}e-h, the development of the concentration field is dissimilar to that of the high injection rate case as the top finger becomes a preferable path instead of the bottom one. Note that the dissolution and precipitation are shown in Figure \ref{fig:c_prog_high_rate_hmc_2d_l3_cag} are a combined effect of injecting water that is unsaturated with respect to calcite and fluid pressure changes. It is clear that the majority of the dissolution occurs due to the transport of the injected water in the porous domain. For the animated version of Figure \ref{fig:c_prog_high_rate_hmc_2d_l3_cag}, please refer to Videos \href{https://figshare.com/s/8a26a94fb3aa1e920433}{1} and \href{https://figshare.com/s/8a26a94fb3aa1e920433}{2}. These videos represent the flow and concentration field as well as $\frac{\partial\phi_c}{\partial t}$ and illustrate the applicability of the presented coupled model for heterogeneous porous media. 

Importantly, this example has illustrated the capability of our proposed method---which is equipped with the EG method---for handling discontinuous material properties across different layers and the sharp interface of the concentration species. Moreover, we have again observed the expected physical and chemical phenomena, including solid deformation, viscous fingering, and dissolution/precipitation.

\subsection{Example 3}\label{ex:3}

In the given computational domain $\Omega_{500}$ = $[0, 100] \times [0, 30]$, we investigate the setup with the heterogeneous 
$\bm{k}$  values as shown in Figure \ref{fig:hmc_2d_cag}c.
A random field generator \cite{nick2009modeling} is utilized to generate a heterogeneous permeability field with a given mean permeability of $\bm{k} = 1 \times 10^{-10} \mathbf{I}$ \si{m^2}, variance of 0.5, and correlation lengths in $x$- and $y$-direction of 5 and 1 m, respectively. 
The heterogeneous permeability field varies in two orders of magnitude. 
All other physical parameters are the same as in the previous examples.

\begin{figure}[!ht]
   \centering
        \includegraphics[width=16.0cm, keepaspectratio]{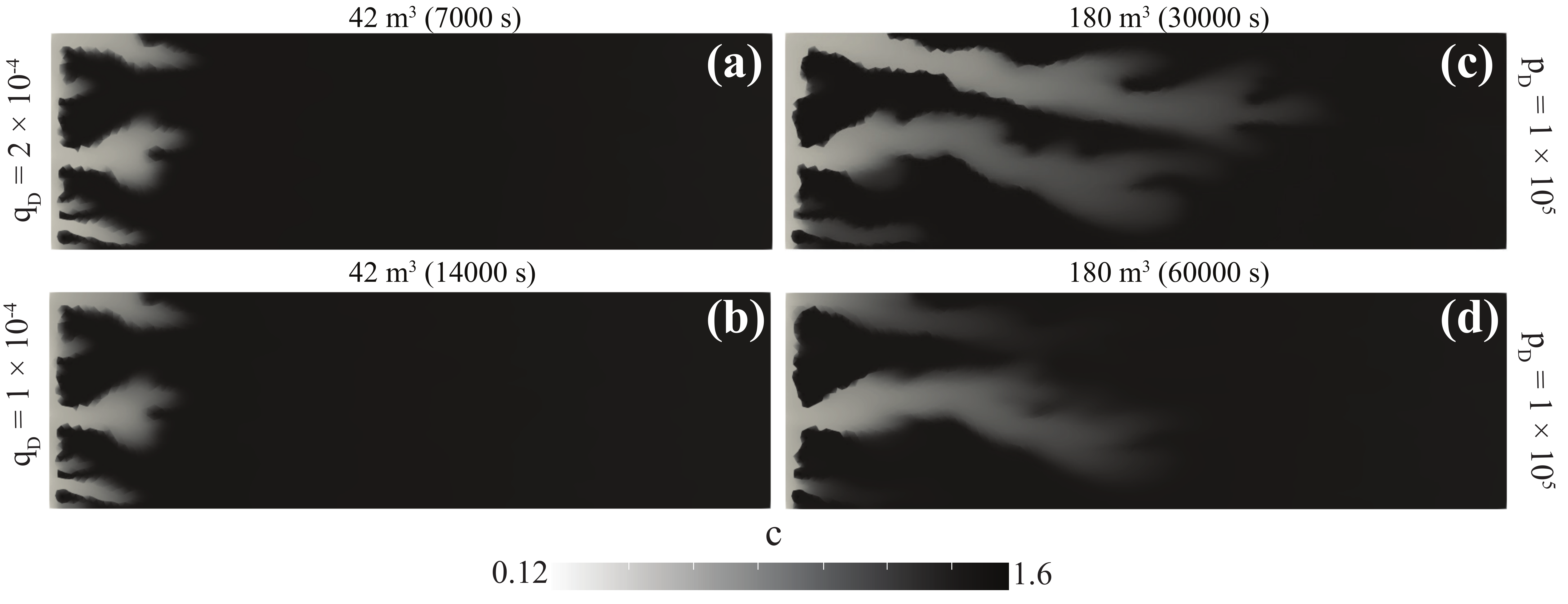}
   \caption{Example 3: concentration fields, $c$; $\mathrm{I.V.} = 42$ \si{m^3} using ($\bm{\mathrm{a}}$) $q_D = 2 \times 10^{-4}$ \si{m/s} at $\partial \Omega_{1}$ and ($\bm{\mathrm{b}}$) $q_D = 1 \times 10^{-4}$ \si{m/s} at $\partial \Omega_{1}$; and $\mathrm{I.V.} = 180$ \si{m^3} using ($\bm{\mathrm{c}}$) $q_D = 2 \times 10^{-4}$ \si{m/s} at $\partial \Omega_{1}$ and ($\bm{\mathrm{d}}$) $q_D = 1 \times 10^{-4}$ \si{m/s} at $\partial \Omega_{1}$. The boundary conditions shown in this picture are corresponding to $\partial \Omega_{1}$ and $\partial \Omega_{3}$ of the mass balance equation \eqref{eq:mass_balance}.}
   \label{fig:c_step1_hmc_2d_het_cag}
\end{figure}

Here, we focus on the interplay between the heterogeneous permeability and the HMC coupled processes. 
Similar to the previous examples, two different injection rates are applied.
In Figure \ref{fig:c_step1_hmc_2d_het_cag}, the concentration fields are illustrated for two different injection rates at different injected fluid volumes ($\mathrm{I.V.} = 42$ \si{m^3} and $\mathrm{I.V.} = 180$ \si{m^3}). 
Unlike the two previous examples, the preferential paths are established not only because of the flow instability resulting from the $\mu$ difference but also due to the high $\bm{k}$ channels inherited from the nature of heterogeneous porous media. During the early time, the concentration field of the high, $q_D = 2 \times 10^{-4}$ \si{m/s}, and low, $q_D = 1 \times 10^{-4}$ \si{m/s}, injection rate cases are similar, see Figure \ref{fig:c_step1_hmc_2d_het_cag}a-b. 
The results of the concentration with the effects from the reaction are different at a later time (see Figure \ref{fig:c_step1_hmc_2d_het_cag}c-d). During the early time for both cases, the developed fingers follow the high permeable paths. At a later time, however, the results of the two scenarios are very different. For the high injection rate case, the top finger continues developing while the middle and the bottom fingers disappear. The result of the low injection rate case, however, shows that the top and the bottom fingers perish while the middle finger progresses. 

\begin{figure}[!ht]
   \centering
        \includegraphics[width=16.0cm, keepaspectratio]{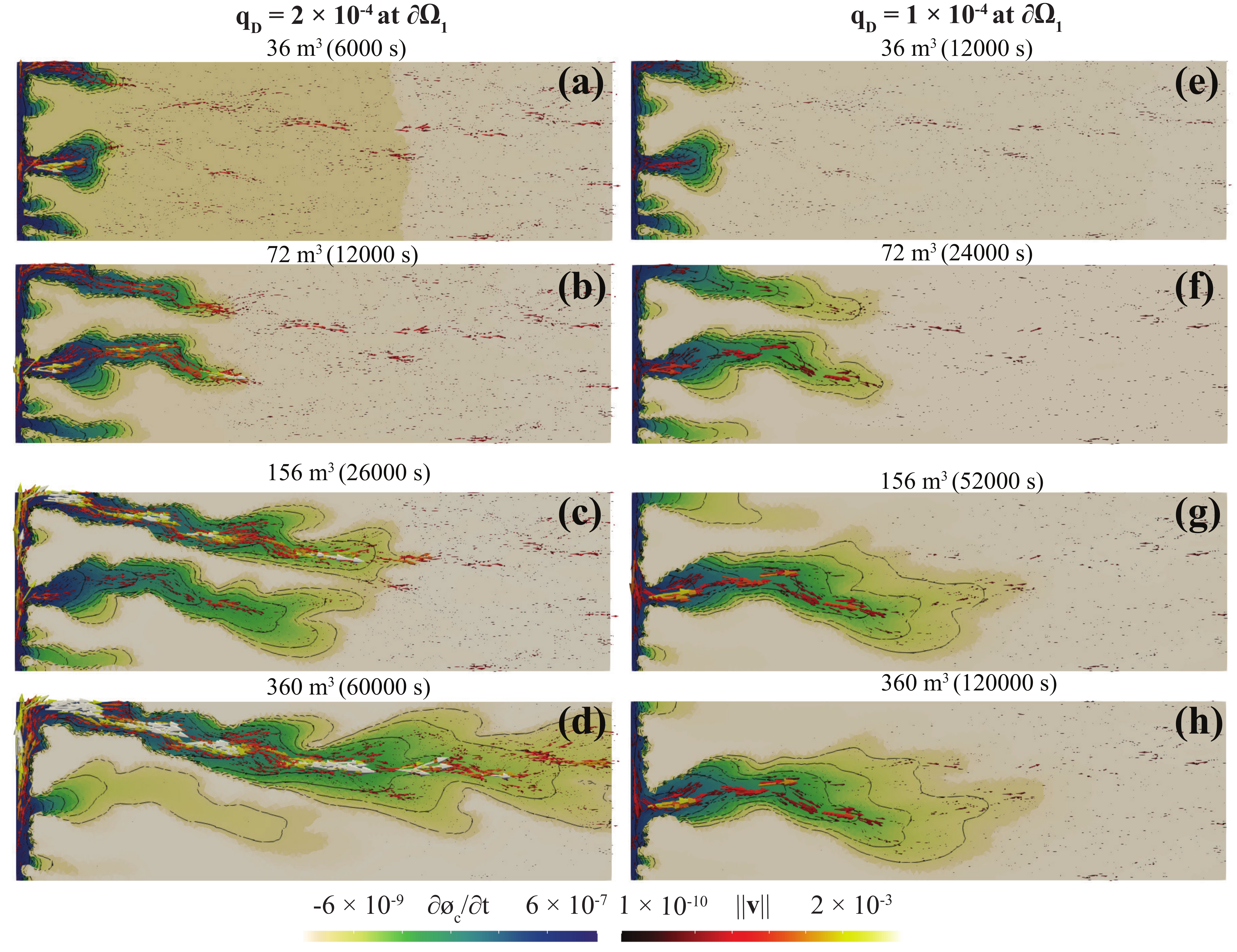}
   \caption{Example 3: the results of the rate of change of porosity according to calcite dissolution/precipitation ($\frac{\partial\phi_c}{\partial t}$) shown in surface plot, concentration ($c$), shown in black contour (10 contours ranging from 0.12 to 1.6), and the fluid velocity ($\bm{q}$) shown in arrows with $q_D = 2 \times 10^{-4}$ \si{m/s} at $\partial \Omega_{1}$; ($\bm{\mathrm{a}}$) $\mathrm{I.V.} = 36$ \si{m^3} ($t = 6000$ \si{s}), ($\bm{\mathrm{b}}$) $\mathrm{I.V.} = 72$ \si{m^3} ($t = 12000$ \si{s}), ($\bm{\mathrm{c}}$) $\mathrm{I.V.} = 156$ \si{m^3} ($t = 26000$ \si{s}), and ($\bm{\mathrm{d}}$) $\mathrm{I.V.} = 360$ \si{m^3} ($t = 60000$ \si{s}), and with $q_D = 1 \times 10^{-4}$ \si{m/s} at $\partial \Omega_{1}$; ($\bm{\mathrm{e}}$) $\mathrm{I.V.} = 36$ \si{m^3} ($t = 1200$ \si{s}), ($\bm{\mathrm{f}}$) $\mathrm{I.V.} = 72$ \si{m^3} ($t = 24000$ \si{s}), ($\bm{\mathrm{g}}$) $\mathrm{I.V.} = 156$ \si{m^3} ($t = 32000$ \si{s}), and ($\bm{\mathrm{h}}$) $\mathrm{I.V.} = 360$ \si{m^3} ($t = 120000$ \si{s}).}
   \label{fig:c_prog_high_rate_hmc_2d_het_cag}
\end{figure}

Figure \ref{fig:c_prog_high_rate_hmc_2d_het_cag} provides further insight into the reactive flow dynamics. It shows for both injection scenarios how the reaction fronts and flow fields evolve in time. As mentioned earlier, all the initial fingers at the beginning vanish except one that reaches the outlet $\partial \Omega_{3}$. The flow velocity field variations in time depict the emergence of the dominant finger. Note that the magnitude of the mechanical deformation is higher than that of the calcite dissolution/precipitation and similar to what was observed in example 1. Therefore changes in porosity due to chemical reaction have a second-order effect on permeability compared to that of induced by the mechanical deformation. Videos \href{https://figshare.com/s/8a26a94fb3aa1e920433}{3} and \href{https://figshare.com/s/8a26a94fb3aa1e920433}{4} representing the flow and concentration field as well as $\frac{\partial\phi_c}{\partial t}$ illustrate the applicability of the presented coupled model for heterogeneous porous media.

Next, we investigate the local mass conservation property of the proposed framework in the heterogeneous domain. The local mass conservation of each cell at each time step, $\mathrm{r^n_{ mass }}$, is calculated by
\begin{equation}\label{eq:mass_loss}
\mathrm{r^n}_{\mathrm{mass}}:=\int_{T} \left(\frac{1}{M}+\frac{\alpha^{2}}{K}\right) \frac{p^{n}-p^{n-1}}{\Delta t^n}+\frac{\alpha}{K} \frac{\sigma_{v}^{n}-\sigma_{v}^{n-1}}{\Delta t^n} + \frac{\hat{\phi_c}^{n}-\phi_c^{n-1}}{\Delta t^n} d V+\sum_{e \in \mathcal{E}_{h}} \int_{e} \bar{\bm{q}}^{n} \cdot\left.\mathbf{n}\right|_{\mathrm{e}} d S,
\end{equation}
and the discrete numerical flux approximated by BDM, $\bar{\bm{q}}^{n} \cdot\left.\mathbf{n}\right|_{\mathrm{e}}$, is defined by 
\begin{align}
\bar{\bm{q}}^{n}  &:=\bm{q}_h^{n} \quad \forall T \in \mathcal{T}_h, \label{eq:flux_internal} \\
\bar{\bm{q}}^{n} \cdot\left.\mathbf{n}\right|_{\mathrm{e}}
&:= - q_{D} \quad \forall e \in \mathcal{E}_{h}^{N,m},\label{eq:flux_neuman} \\
\bar{\bm{q}}^{n} \cdot\left.\mathbf{n}\right|_{\mathrm{e}}
&:= - \bm{q}_h^{n} \cdot \mathbf{n} \quad \forall e \in \mathcal{E}_{h}^{D,m}.\label{eq:rf_dir}
\end{align}

\begin{figure}[!ht]
   \centering
        \includegraphics[width=16.0cm, keepaspectratio]{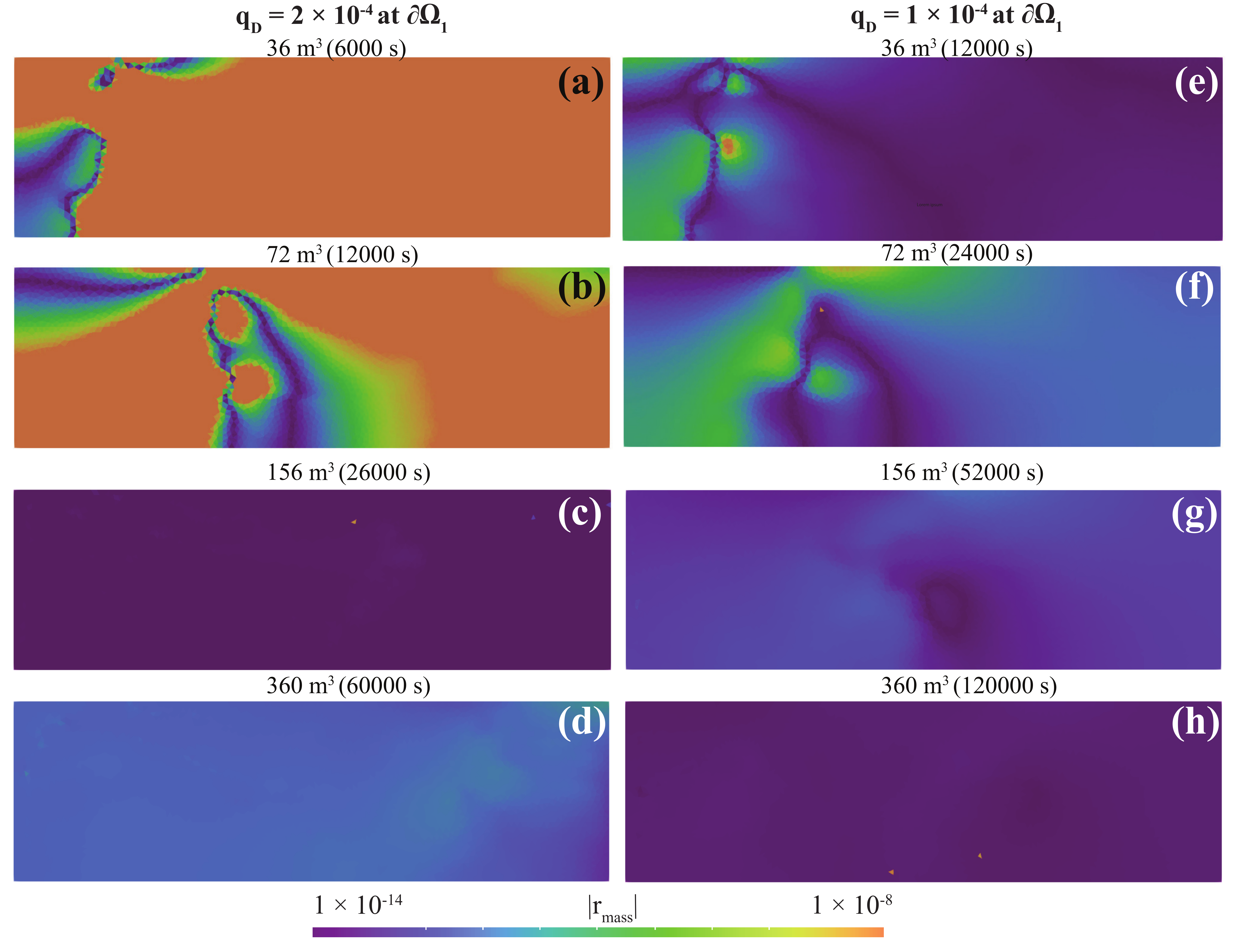}
   \caption{Example 3: the illustration of the local mass conservative property with $q_D = 2 \times 10^{-4}$ \si{m/s} at $\partial \Omega_{1}$; ($\bm{\mathrm{a}}$) $\mathrm{I.V.} = 36$ \si{m^3} ($t = 6000$ \si{s}), ($\bm{\mathrm{b}}$) $\mathrm{I.V.} = 72$ \si{m^3} ($t = 12000$ \si{s}), ($\bm{\mathrm{c}}$) $\mathrm{I.V.} = 156$ \si{m^3} ($t = 26000$ \si{s}), and ($\bm{\mathrm{d}}$) $\mathrm{I.V.} = 360$ \si{m^3} ($t = 60000$ \si{s}), and with $q_D = 1 \times 10^{-4}$ \si{m/s} at $\partial \Omega_{1}$; ($\bm{\mathrm{e}}$) $\mathrm{I.V.} = 36$ \si{m^3} ($t = 1200$ \si{s}), ($\bm{\mathrm{f}}$) $\mathrm{I.V.} = 72$ \si{m^3} ($t = 24000$ \si{s}), ($\bm{\mathrm{g}}$) $\mathrm{I.V.} = 156$ \si{m^3} ($t = 32000$ \si{s}), and ($\bm{\mathrm{h}}$) $\mathrm{I.V.} = 360$ \si{m^3} ($t = 120000$ \si{s}).}
   \label{fig:r_mass_high_rate_hmc_2d_het_cag}
\end{figure}

In Figure \ref{fig:r_mass_high_rate_hmc_2d_het_cag}, 
the values of $\mathrm{r^n_{ mass }}$ are illustrated for each case and time.
One could see that the magnitude of $\mathrm{r^n_{ mass }}$ is always less than $1 \times 10^{-5}$, which is the tolerance set for the fixed-stress loop, see Algorithm \ref{al:1}; therefore, the framework is locally mass conservative. We note that the high injection rate case tends to the higher value of the magnitude of $\mathrm{r^n_{mass}}$ than that of the low injection rate case.

\subsection{Example 4}

Lastly, we investigate the performance of the proposed framework when the permeability field is anisotropic, and the grid is unstructured, as shown in Figure~\ref{fig:hmc_2d_cag}d.
In the computational domain $\Omega_{500}$ = $[0, 100] \times [0, 30]$, see Figure \ref{fig:hmc_2d_cag}a, we consider the anisotropic permeability field to emphasize the capability of our proposed algorithm. The permeability tensor of this example is defined as follows:

\begin{equation}  \label{eq:permeability_matrix_ani}
\bm{k}:=\left[ \begin{array}{cc} k_{xx} & 0.0 \\ 0.0 & 0.1 k_{xx} \end{array}\right],
\end{equation}


\noindent
where $k_{xx} = 8.8 \times 10^{-10}$ \si{m^2} and all other parameters are similar to all other cases.

\begin{figure}[!ht]
   \centering
        \includegraphics[width=16.0cm, keepaspectratio]{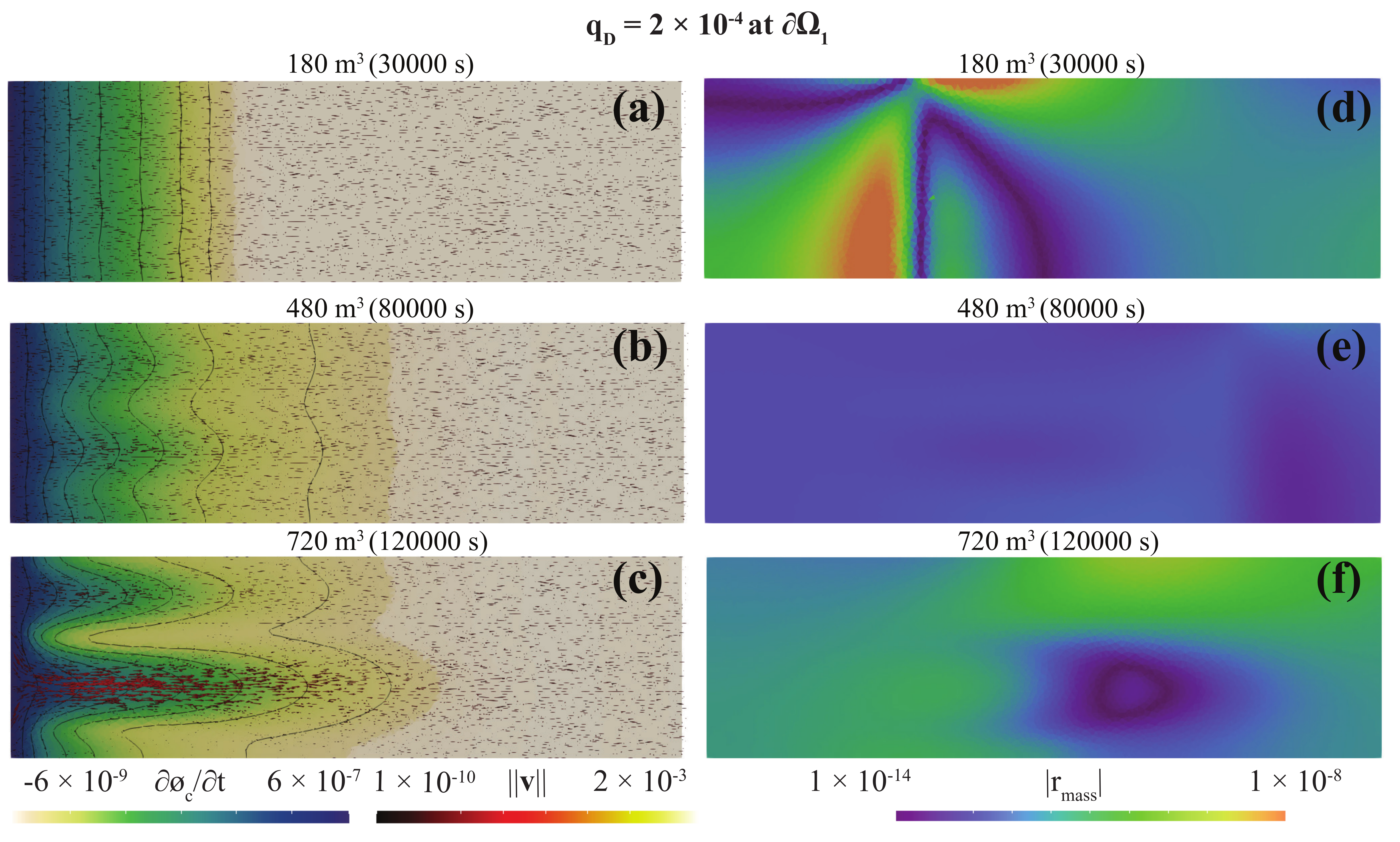}
   \caption{Example 4: the results of the rate of change of porosity according to calcite dissolution/precipitation ($\frac{\partial\phi_c}{\partial t}$) shown in surface plot, concentration ($c$), shown in black contour (10 contours ranging from 0.12 to 1.6), and the fluid velocity ($\bm{q}$) shown in arrows with $q_D = 2 \times 10^{-4}$ \si{m/s} at $\partial \Omega_{1}$; ($\bm{\mathrm{a}}$) $\mathrm{I.V.} = 180$ \si{m^3} ($t = 30000$ \si{s}), ($\bm{\mathrm{b}}$) $\mathrm{I.V.} = 480$ \si{m^3} ($t = 80000$ \si{s}), and ($\bm{\mathrm{c}}$) $\mathrm{I.V.} = 720$ \si{m^3} ($t = 120000$ \si{s}), and the local mass conservative property; ($\bm{\mathrm{d}}$) $\mathrm{I.V.} = 180$ \si{m^3} ($t = 30000$ \si{s}), ($\bm{\mathrm{e}}$) $\mathrm{I.V.} = 480$ \si{m^3} ($t = 80000$ \si{s}), and ($\bm{\mathrm{f}}$) $\mathrm{I.V.} = 720$ \si{m^3} ($t = 120000$ \si{s}).}
   \label{fig:c_prog_high_rate_hmc_2d_ani_cag}
\end{figure}

Figure \ref{fig:c_prog_high_rate_hmc_2d_ani_cag} shows the reactive flow dynamics and the residual of mass. 
We observe that the flow in the horizontal direction dominates the flow in the vertical direction since the permeability in the horizontal direction is ten times higher than that of the vertical direction. Figure \ref{fig:c_prog_high_rate_hmc_2d_ani_cag}d-f illustrate that the proposed framework is locally mass conservative as the residual of mass values are always less than $1 \times 10^{-5}$, which is the tolerance set for the fixed-stress loop.

\subsection{Discussion}

The main observations of the foregoing numerical examples can be summarized as follows: 

\begin{enumerate}
    \item The injection rate supplied at the inlet boundary is critical in defining flow behavior. The preferential flow paths developed through time are significantly different with different injection rates. Besides, the injection flow rate also controls the development of the advection and reaction fronts. 
    \item Using the applied set of the input parameters resulted in a more noticeable mechanical effect on the change in $\phi$ (and subsequently in $\bm{k}$) compared to that of the calcite dissolution/precipitation effect. We note that this observation could vary with different sets of input parameters and required to be further investigated. The change in $\mu$ resulted from the change in $c$ is significant, resulting in the development of preferential flow paths.
    \item The results of both homogeneous and heterogeneous as well as isotropic and anisotropic permeability field show that our framework preserves mass locally. This property is essential for the coupled HMC system.
\end{enumerate}

In terms of computational efficiency, it is noted that the iteration number for the fixed-stress iteration was around three (four for the example 3) at the initial time stage, but it only required two iterations for the rest of the time for all the presented examples.
For all examples, we have 31934, 23818, 7852, 11910 degrees of freedom for the displacement, flux, pressure, and concentration fields, respectively.
The computational time was around $4.78 \times 10^{-5}$ second per degrees of freedom per each time step.  All simulations were computed on XeonE5\_2650v4 with a single thread.

\section{Conclusion}\label{sec:conclusion}

This paper has presented a mixed finite element framework for coupled hydro-mechanical-chemical processes in heterogeneous porous media.
The main advantage of the proposed framework is its relatively affordable cost to attain local conservation regardless of material anisotropy, thanks particularly to the use of the EG method. Through several numerical examples, we have demonstrated the performance and capabilities of the proposed framework with a focus on local conservation. 
The numerical results have highlighted how the overall behavior is influenced by different processes, including solid deformation, calcite dissolution, and fluid viscosity alteration.  
The developed numerical model can provide insight into how the interactions among HMC processes and heterogeneity manifest themselves at a larger scale.
Future work includes an extension of the modeling framework to coupled thermo-hydro-mechanical-chemical processes in heterogeneous and/or fractured porous media. 

\section{Acknowledgements}

This research has received financial support from the Danish Hydrocarbon Research and Technology Centre under the Advanced Water Flooding program. 
The computational results in this work have been produced by the multiphenics library~\cite{Ballarin2019}, which is an extension of FEniCS \cite{AlnaesBlechta2015a} for multiphysics problems. We acknowledge the developers of and contributors to these libraries. 
TK also thanks the 2019 Computers \& Geosciences Research grant for the additional support.
SL is supported by the National Science Foundation under Grant No. NSF DMS-1913016.
FB thanks Horizon 2020 Program for Grant H2020 ERC CoG 2015 AROMA-CFD project 681447 that supported the development of multiphenics.
JC acknowledges support from the Research Grants Council of Hong Kong (Project 27205918).

\section{CRediT authorship contribution statement} \label{sec:credit}

\textbf{T. Kadeethum}: Conceptualization, Formal analysis, Software, Validation, Writing - original draft, Writing - review \& editing. \textbf{S. Lee}: Conceptualization, Formal analysis, Supervision, Validation, Writing - review \& editing. \textbf{F. Ballarin}: Conceptualization, Formal analysis, Software, Supervision, Validation, Writing - review \& editing. \textbf{J. Choo}: Conceptualization, Formal analysis, Supervision, Writing - review \& editing. \textbf{H.M. Nick}: Conceptualization, Funding acquisition, Supervision, Writing - review \& editing.

\section{Computer code availability}

The scripts used to produce these results are available at this \href{https://github.com/teeratornk/supplementary_scripts_for_HMC_manuscript}{Git} repository. The main dependencies are Numpy ($\ge$ 1.16.5), \href{https://fenicsproject.org/}{FEniCS} ($\ge$ 2018.1.0) with  \href{https://www.mcs.anl.gov/petsc/}{PETSc} ($\ge$ 3.10.5) and petsc4py ($\ge$ 3.10), and \href{https://mathlab.sissa.it/multiphenics}{multiphenics} ($\ge$ 0.2.0).

\biboptions{sort&compress,square,comma,numbers}
\bibliographystyle{elsarticle-num}

\bibliography{cite.bib}

\end{document}